\newcommand{\beq}[1]{\begin{equation}\label{#1}}
\newcommand{\eeq}{\end{equation}}
\newcommand{\beqar}[1]{\begin{eqnarray}\label{#1}}
\newcommand{\eeqar}{\end{eqnarray}}
\newcommand{\Pslash}{ P \mskip -11mu / \mskip 4mu}
\newcommand{\qslash}{ q \mskip  -8mu / \mskip 1mu}
\documentstyle[12pt,dina4,preprint,aps,epsf]{revtex}
\begin{document}
\draft
\title{OPE analysis of the nucleon scattering tensor including
weak interaction and finite mass effects}
\author{
        M. Maul, B. Ehrnsperger, E. Stein, and A. Sch\"afer}
\address{
%\begin{minipage}{14cm}
%\flushleft
%\vspace{1cm}
        Institut f\"ur Theoretische Physik, J.~W.~Goethe Universit\"at Frankfurt\\
        $^{~}$Postfach 11 19 32, 60054 Frankfurt am Main, Germany\\
%\end{minipage}}
}
\maketitle
\begin{abstract}
We perform a systematic operator product expansion of the most general form of 
the nucleon scattering tensor $W_{\mu \nu}$ including electro-magnetic
and weak interaction processes.
Finite quark masses are taken into account and a number of higher-twist corrections 
are included.
In this way we derive relations between  the lowest moments of
all 14 structure functions and matrix elements of local operators.
Besides reproducing well-known results, new sum rules 
for parity-violating polarized structure functions and new mass correction terms
are obtained.
\end{abstract}
\vspace*{5cm}  
\hfill UFTP preprint 404/1995
\pagebreak
%%%%%%%%%%%%
%\psdraft
%%%%%%%%%%%
\section{Introduction}
\noindent
Deep inelastic lepton-nucleon scattering (DIS) provides the cleanest method
to understand the internal structure of the nucleon \cite{Rob90}.
Presently such measurements performed at HERA, CERN, SLAC, 
and other accelerators  
concentrate on electro-magnetic DIS. Theoretically 
weak interaction processes can provide valuable additional information 
\cite{Ji93}, allowing e.g. a flavor decomposition of structure functions. 
Such experiments are extremely difficult and require careful and detailed
analysis, but the possible scientific progress justifies a serious
consideration of them \cite{An94}.

All DIS experiments aim at measuring the structure functions of the nucleon.
In the framework of operator product expansion (OPE),
moments of structure functions can be related to nucleon 
matrix elements of well defined local operators 
which have to be calculated nonperturbatively, e.g. in lattice QCD
\cite{Horsley}.
In some cases those matrix elements occur in other strong 
interaction processes and sum rules can be derived.
The polarized Bjorken sum rule \cite{Bj66}, relating nucleonic $\beta$-decay and the 
first moment of the polarized spin structure function $\int dx g_1(x)$,
is an especially interesting case.
Such sum rules offer a convincing way of testing perturbative 
and non perturbative QCD.

With the accuracy of QCD tests reaching the per cent or even per mille
level, the higher-twist contributions to DIS become relevant.
While quite a number of theoretical investigations on their form exists
scattered in the literature, no systematic presentation of all
contributions up to the second order in the expansion of the fermion
propagator within the formalism of OPE has been given so far. As compared
to the interactionless quark parton model used in the analysis of \cite{An94}, 
the OPE formalism is of general validity and includes the QCD interactions.
This can be seen in the fact that within the framework of the quark parton
model \cite{An94} obtains for example $g_2^\gamma(x)=0$, while the OPE
predicts $\int_0^1 dx x^2 g_2^\gamma (x) $ being different from zero.

This paper gives a complete determination of the lowest moments
of the 14 nucleon structure functions which parametrize the most
general form of the nucleon scattering tensor (only restricted
by Lorentz and time-reversal invariance).
The paper is organized as follows:
In section II we define all relevant quantities and follow the standard 
procedure of OPE to isolate contributions of well defined twist. 
This section is supplemented by appendix A in which the decomposition of all
relevant operators is performed explicitly. Using these results we analyze 
$\gamma$ and $Z^0$ exchange in section III and $W^{\pm}$ exchange in section IV.
Our final results are presented in section V and the conclusions in section VI.
A discussion of the contributions from cat-ear  diagrams is delegated to 
appendix B.

Our main results are new 
higher-twist and quark mass corrections to the parity violating
polarized structure functions.

\section{Definitions and Spin Decomposition}
\noindent
DIS processes 
can be distinguished according to the exchanged boson,
namely $\gamma$, $Z^0$, or $W^{\pm}$. 
The cross sections for these processes are
\begin{eqnarray}
\label{E1}
\frac{ d^2 \sigma ^{(\gamma)}}{dE'd \Omega} &=& 
\frac{E'}{EM} \frac{\alpha^2}{(Q^2)^2}
L_{\mu \nu}^{(\gamma)} W^{(\gamma)\mu \nu} \quad,
\\
\nonumber \\
\frac{ d^2 \sigma ^{(Z^0)}}{dE'd \Omega} &=&
\frac{E'}{EM} \frac{\alpha^2}{(Q^2+(M_{Z^0})^2)^2}
L_{\tau \sigma}^{(Z^0)} W^{(Z^0)}_{\mu \nu} 
\left(- g^{\mu \tau}   + \frac{q^\mu q^\tau}{M^2_{Z^{0}}} \right)
\left(- g^{\nu \sigma} + \frac{q^\nu q^\sigma}{M^2_{Z^{0}}} \right) \quad,
\\
\nonumber \\
\frac{ d^2 \sigma ^{(W^{\pm})}}{dE'd \Omega} &=&     
\frac{E'}{EM} \frac{\alpha^2}{(Q^2+(M_{W^{\pm}})^2)^2}
L_{\tau \sigma}^{(W^{\pm})} W^{(W^{\pm})}_{\mu \nu}      
\left(- g^{\mu \tau}   + \frac{q^\mu q^\tau}{M^2_{W^{\pm}}} \right)
\left(- g^{\nu \sigma} + \frac{q^\nu q^\sigma}{M^2_{W^{\pm}}} \right) \quad .
\end{eqnarray}
Here $E$ is the energy of the incoming lepton,
$E'$ the energy of the outgoing lepton, and $M$ is the nucleon mass. 
$\Omega$ is the solid angle of the outgoing lepton.
As usual $q$ is the photon, respectively boson momentum 
with $Q^2 =-q^2$,
$\alpha$ is the electromagnetic coupling constant. 
The weak interaction coupling constants are absorbed into the 
lepton ($L_{\mu \nu}$) and nucleon ($W_{\mu \nu}$) scattering tensors.

These tensors can be 
written in the following form:
\begin{eqnarray}
L_{\mu \nu}^{(Pr)} & = &  \sum_{\sigma'} 
\langle k \sigma  |\left( j_\mu^{(L,Pr)}(0)\right)^\dagger |k' \sigma' \rangle
\langle k'\sigma' | j_\nu^{(L,Pr)}(0) |k  \sigma  \rangle \quad,
\nonumber \\
\left(W_{\mu \nu}^{(Pr)}\right)_\lambda  & = & \frac{1}{2\pi} \sum_X 
(2 \pi)^4\delta^4(p_X-p-q) \frac{1}{2} 
\langle p \lambda |\left(j_\mu^{(H,Pr)}(0)\right)^\dagger | X \rangle
\langle X|j_\nu^{(H,Pr)}(0) |p \lambda  \rangle 
\nonumber \\
 & = & \frac{1}{4\pi} \int d^4 \xi e^{iq\xi}
\langle p \lambda |\left[
\left( j_\mu^{(H,Pr)}(\xi) \right)^\dagger, 
j_\nu^{(H,Pr)}(0)\right]_- |p \lambda  \rangle 
\quad.
\end{eqnarray}
The index $Pr = \gamma, Z^0, W^+, W^-$
indicates the process under consideration. The lepton current is 
labeled by $L$ and the hadronic one by $H$.
$k^\mu$ is the four-momentum of the incoming lepton and $\sigma^\mu$ its spin.
${k'}^\mu$ and ${\sigma'}^\mu$ are four-momentum and spin of the outgoing
lepton. The nucleon four-momentum is $p^\mu$, ($p^2 = M^2$) 
and $\lambda = \pm 1/2$ denotes its polarization (cf. figure 1).
Let $ S_\mu$ be its spin vector , then according to \cite{Man}
\begin{equation}
\label{2}
W_{\mu \nu}^{(Pr)} (q,p,S) = \sum_\lambda \langle \lambda |\hat \rho| \lambda \rangle
                   \left( W_{\mu \nu}^{(Pr)}(q,p)\right)_\lambda \quad,
\end{equation}
with the spin density matrix 
\begin{equation}
\hat \rho := \frac{1}{2} \left( 1 + \vec \sigma \cdot \frac{\vec S}{M} \right)
\quad,
\end{equation}
and $S^\mu = (0,\vec S)$ in the 
rest frame of the target ($S^2= -M^2$, $S\cdot p = 0$).
We use covariant normalization for the lepton states
$\langle k \sigma| k' \sigma' \rangle = 2k_0 (2\pi)^3 \delta(\vec k - \vec k')
\delta_{\sigma,\sigma'}$, and
$\langle pS|p'S' \rangle = 2p_0 (2\pi)^3\delta(\vec p - \vec p')
\delta_{S,S'}$ for the nucleon states in $W_{\mu \nu}^{(PR)}(q,p,S)$.
The lepton currents read
\begin{eqnarray}
j_\mu^{(L,\gamma)}(\xi) &=& \sum_e \overline e(\xi) \gamma_\mu e(\xi) \quad,
\nonumber \\
j_\mu^{(L, Z^0)}(\xi)  &=&  \sum_l \overline l (\xi) \gamma_\mu  
                           (V_l' + A_l'\gamma_5) l(\xi) \quad, 
\nonumber \\
j_\mu^{(L,W^+)}(\xi)    &=& \sum_{e,\nu} \overline e (\xi) \gamma_\mu 
                           ({V'}^+  +{A'}^+\gamma_5) \nu (\xi)
                           \quad,
\nonumber \\
j_\mu^{(L,W^-)}(\xi)    &=& \sum_{e,\nu} \overline \nu (\xi) \gamma_\mu
                           ({V'}^-  +{A'}^- \gamma_5) e (\xi)
                           \quad,
\end{eqnarray} 
where $l \in \{ e^-, \mu^-, \tau^-,\nu_e, \nu_\mu, \nu_\tau\}$, 
$ e \in \{ e^-, \mu^-, \tau^-\}$ and $ \nu \in \{ \nu_e, \nu_\mu, \nu_\tau\}$. 
The nucleon current is written as 
\begin{eqnarray}
\label{hadcur}
j_\mu^{(H,\gamma)} (\xi) &=& \sum_f Q_f \overline q_f(\xi) \gamma_\mu  q_f(\xi)
\quad,
\nonumber \\
j_\mu^{(H,Z^0)} (\xi) &=& \sum_f \overline q_f(\xi) \gamma_\mu  
(V_f + A_f \gamma_5) q_f(\xi) \quad,
\nonumber \\
j_\mu^{(H,W^+)} (\xi) &=& \overline q_u(\xi) \gamma_\mu  
(V^+_{ud} + A^+_{ud} \gamma_5) q_d(\xi) + {\rm further\; flavor\; combinations} \quad,
\nonumber \\
j_\mu^{(H,W^-)} (\xi) &=& \overline q_d(\xi) \gamma_\mu           
(V^-_{du} + A^-_{du} \gamma_5) q_u(\xi) + {\rm further\; flavor\; combinations} \quad.
\end{eqnarray}
The constants $V$ and $A$ follow from the standard model (see for example
\cite{MS91}).
$Q_f$ is the electromagnetic quark charge and 
$f$ runs over all flavors.
All quark states considered here are mass eigenstates, 
i.e. Cabbibo-Kobayashi-Maskawa mixing 
\cite{KM73} is suppressed for the sake of a compact notation.
To simplify notation we
suppress the obvious flavor indices $V^{\pm}_{ud} \rightarrow V^{\pm}$ in the
following. 

A slight modification arises for the  $\gamma$ and $Z^0$
interference terms.
\begin{eqnarray}
\frac{ d^2 \sigma ^{({\rm int.}\; \gamma,Z^0)}}{dE'd \Omega} &=&
\frac{E'}{EM} \frac{\alpha^2}{(Q^2+(M_{Z^0})^2)Q^2}
\left(- g^{\mu \tau} \right)
\left(- g^{\nu \sigma} + \frac{q^\nu q^\sigma}{M^2_{Z^{0}}} \right) 
\nonumber \\
\nonumber \\
&& \times
\left(
L_{\tau \sigma}^{({\rm int.}\; \gamma,Z^0)(1)} 
W_{\mu \nu}    ^{({\rm int.}\; \gamma,Z^0)(1)}
+
L_{\sigma \tau}^{({\rm int.}\; \gamma,Z^0)(2)}
W_{\nu     \mu}^{({\rm int.}\; \gamma,Z^0)(2)} \right) \quad,
\nonumber \\
 \left(W_{\mu \nu}^{({\rm int.}\; \gamma,Z^0)(1)}\right)_\lambda 
 & = & \frac{1}{4\pi} \int d^4 \xi e^{iq\xi}
\langle p \lambda |
\left[
\left( j_\mu^{(H,\gamma)}(\xi) \right)^\dagger, j_\nu^{(H,Z^0)}(0)\right]_-
|p \lambda \rangle \quad,
\nonumber \\
\left(W_{\mu \nu}^{({\rm int.}\; \gamma,Z^0)(2)}\right)_\lambda  
&=&  \frac{1}{4\pi} \int d^4 \xi e^{iq\xi}
\langle p \lambda |
\left[
\left( j_\mu^{(H,Z^0)}(\xi) \right)^\dagger, j_\nu^{(H,\gamma)}(0)\right]_-
|p \lambda \rangle \quad,
\nonumber \\
L_{\mu \nu}^{({\rm int.}\; \gamma,Z^0)(1)}
&=&
\sum_{\sigma'}
\langle k \sigma  |\left( j_\mu^{(L,\gamma)}(0)\right)^\dagger |k' \sigma' \rangle
\langle k'\sigma' | j_\nu^{(L,Z^0)}(0) |k  \sigma  \rangle \quad,
\nonumber \\
L_{\mu \nu}^{({\rm int.}\; \gamma,Z^0)(2)}  
&=&  \sum_{\sigma'}
\langle k \sigma  |\left( j_\mu^{(L,Z^0)}(0)\right)^\dagger |k' \sigma' \rangle
\langle k'\sigma' | j_\nu^{(L,\gamma)}(0) |k  \sigma  \rangle  \quad,
\nonumber \\
\left(W_{\mu \nu}^{({\rm int.}\; \gamma,Z^0)}\right)_\lambda
&=&
\left(W_{\mu \nu}^{({\rm int.}\; \gamma,Z^0)(1)}\right)_\lambda +
\left(W_{\mu \nu}^{({\rm int.}\; \gamma,Z^0)(2)}\right)_\lambda \quad,
\nonumber \\
L_{\mu \nu}^{({\rm int.}\; \gamma,Z^0)}
&=&
L_{\mu \nu}^{({\rm int.}\; \gamma,Z^0)(1)}
+L_{\mu \nu}^{({\rm int.}\; \gamma,Z^0)(2)} \quad.
\end{eqnarray}
Using Lorentz covariance and time-reversal invariance 
the nucleon scattering tensor $W^{(Pr)}_{\mu\nu}(q,P,S)$
can be expressed in terms of 14 structure functions \cite{Ji93}
(In our sign convention all $\epsilon$ structures are multiplied 
by $-i$.):
\begin{eqnarray}
W^{\mu\nu(Pr)}(q,p,S) &=& \left(-g^{\mu \nu}+\frac{q^\mu q^\nu}{q^2} \right)
                      F_1^{(Pr)}(x,Q^2) 
                     + \hat p^\mu \hat p^\nu \frac{1}{\nu}F_2^{(Pr)}(x,Q^2)
\nonumber \\
                  && + q^\mu q^\nu \frac{1}{\nu} F_4^{(Pr)}(x,Q^2)
                      + \left(p^\mu q^\nu + p^\nu q^\mu\right) \frac{1}{2\nu}F_5^{(Pr)}(x,Q^2)
\nonumber \\
                  && -i \epsilon^{\mu \nu \alpha \beta }\frac{q_\alpha}{\nu}
                      \left[S_\beta g_1^{(Pr)}(x,Q^2) + 
             \left( S_\beta - p_\beta \frac{Sq}{\nu} \right) g_2^{(Pr)}(x,Q^2) \right]
\nonumber \\      && -i \epsilon^{\mu \nu \alpha \beta} \left[
                      \frac{p_\alpha S_\beta}{\nu} g_3^{(Pr)}(x,Q^2)  
                     +\frac{q_\alpha p_\beta}{2\nu} F_3^{(Pr)}(x,Q^2) \right]
\nonumber \\      && + \left(-g^{\mu \nu}+\frac{q^\mu q^\nu}{q^2} \right)
                       \frac{Sq}{\nu} a_1^{(Pr)}(x,Q^2)
                     + \hat p^\mu \hat p^\nu \frac{Sq}{\nu^2}a_2^{(Pr)}(x,Q^2)
\nonumber \\
                  && + q^\mu q^\nu \frac{Sq}{\nu^2} a_4^{(Pr)}(x,Q^2)
                  + \left(p^\mu q^\nu + p^\nu q^\mu\right) \frac{Sq}{2\nu^2}a_5^{(Pr)}(x,Q^2)
\nonumber \\
               && + \left(\hat S^\mu \hat p^\nu + \hat S^\nu \hat p^\mu\right) 
                     \frac{1}{2\nu}b_1^{(Pr)}(x,Q^2) 
                  + \left(\hat S^\mu p^\nu + \hat S^\nu p^\mu \right) 
                      \frac{1}{2\nu}b_2^{(Pr)}(x,Q^2).
\end{eqnarray}
Here $\nu = p\cdot q$ and $x = Q^2/(2\nu)$. Furthermore the 
vectors $\hat p_\mu = p_\mu  + q_\mu \, \nu/Q^2$,
and $\hat S_\mu = S_\mu + q_\mu \,S\cdot q/Q^2$, which  
vanish if contracted with
$q_\mu$, are introduced.
Note that terms proportional to $q^\mu$ or $q^\nu$ give contributions
of order $m_l/E$ when contracted with the lepton tensor $L_{\mu\nu}$.
The decomposition yields seven structure functions for 
parity-conserving processes and additional seven for  
parity-non-conserving processes.
All 14 are given in table \ref{tab1}.
Six structure functions 
($F_4,F_5,g_3,a_4,a_5,b_2$) are related to 
non-conserved currents and are non-zero only from non-vanishing quark masses.

In the framework of OPE the moments of structure functions 
can be related to nucleon matrix elements of local operators 
with well defined twist.
To this end 
the nucleon scattering tensor $W^{(Pr)}_{\mu \nu}$ is related
to the virtual forward Compton scattering amplitude $T^{(Pr)}_{\mu\nu}$
\begin{equation}
\label{9}
W^{(Pr)}_{\mu \nu} = \frac{1}{4 \pi i} 
\left(T^{(Pr)}_{\mu \nu}(q_0 + i \epsilon)
-T^{(Pr)}_{\mu \nu}(q_0 - i \epsilon) \right)  \quad,                 
\end{equation}

\begin{equation}
T^{(Pr)}_{\mu \nu}(q_0 \pm i\epsilon)
= \frac{(2\pi)^3 \delta^3(\vec p_X-\vec p -\vec q\;)}
        {{p_X}_0 - p_0  - (q_0 \pm i \epsilon)}
        \sum_X \langle pS| \left(j^{(Pr)}_\mu \right)^\dagger (0) |X\rangle
        \langle X | j^{(Pr)}_\nu(0) |pS \rangle \quad,
\end{equation}
\begin{eqnarray}
\label{Eq1} 
\int_0^1 dx  x^{n}  W^{(Pr)}_{\mu \nu}(x)
= \frac{1}{8\pi i}\oint
\frac{ T^{(Pr)}_{\mu \nu}
(\omega) d\omega}{\omega^{n+2}} \quad,
\end{eqnarray}
with $\omega = 1/x = 2\nu/Q^2$.
In the following we deal with crossing-even and crossing-odd
amplitudes 
\begin{eqnarray}
\label{10}
 T^{(\gamma)}_{\mu \nu}(-q) &=& T^{(\gamma)}_{\nu \mu}(q) \quad,
\nonumber \\
 T^{(Z^0)}_{\mu \nu}(-q) &=& T^{(Z^0)}_{\nu \mu}(q) \quad,
\nonumber \\
T_{\mu \nu} ^{(\overline \nu + \nu)}(-q) 
&=& T_{\nu \mu} ^{(\overline \nu + \nu)}(q) \quad,
\nonumber \\
T_{\mu \nu} ^{(\overline \nu - \nu)}(-q) 
&=& - T_{\nu \mu} ^{(\overline \nu - \nu)}(q) \quad,
\end{eqnarray}
where $T_{\mu \nu} ^{(\overline \nu \pm \nu)}
= T^{(W^-)}_{\mu \nu} \pm T^{(W^+)}_{\mu \nu}$.
Note that the formal equation (\ref{Eq1}) is only valid for certain
moments of the structure functions because its derivation requires for closing
the integration contour according
to fig. 2 for the $n$-th moment  
\begin{equation}
\label{ntemoments}
T^{(Pr)}_{\mu\nu}(\omega) = 
- (-)^n \; T^{(Pr)}_{\mu\nu}(-\omega) \quad.
\end{equation}
This condition if fulfilled in the case of 
the crossing-even  amplitudes such as $ T^{(\gamma)}_{\mu \nu}$, 
$ T^{(Z_0)}_{\mu \nu}$ and $T_{\mu \nu} ^{(\overline \nu + \nu)}$
for the odd moments (even $n$) of the structure functions
$F_2$, $F_4$, $a_2$, $a_4$, $b_1$, $b_2$, $g_1$, $g_2$ and $F_3$
and for the even moments of 
$F_1$, $F_5$, $a_1$, $a_5$, and $g_3$.
For the crossing-odd amplitude $T_{\mu \nu} ^{(\overline \nu - \nu)}$
the situation is just inverted.
Furthermore performing the integration in  
equation (\ref{Eq1}) is possible only if $T_{\mu \nu}(\omega)$ vanishes
fast enough for $\omega \rightarrow \infty$, i.e, when the integration
over the circle in infinity in figure 2 does not contribute. This
is true for $n \geq 1$ but not necessarily for $n=0$.
(In practice one expects nonsinglet sum rules to converge
better since there is no pomeron contribution and nonsinglet
structure functions are less singular for small $x$.) 
%since the $1/x \to \infty$ behavior cannot be related to
%the pomeron intercept.
\\ \\
To analyze the Compton forward scattering amplitude 
\begin{equation}
\label{7}
T^{(Pr)}_{\mu \nu}(q,p,S)
 =  i \int d^4 \xi e^{i q \xi} \langle p S| T\left((j^{(Pr)}_\mu)^\dagger (\xi)
 j^{(Pr)}_\nu(0)\right)|p S \rangle 
\end{equation}
we insert the nucleon currents Eq.(\ref{hadcur}) 
which we now write in the general form
\begin{equation}
j_{\mu ff'}(\xi) = \overline q_f( \xi) \Gamma_{\mu ff'} q_{f'}( \xi)\quad .
\end{equation}
Note that $\Gamma_{\mu ff'} = \Gamma_{\mu f'f}$.
Restricting ourselves for the moment to the handbag diagram 
(cf. figure 3)
we write
\begin{eqnarray}
\label{19}
T_{\mu \nu f f'} 
&=& i \int d^4 \xi e^{iq\xi}
   \langle pS| \left\{ \overline q_f (\xi)
           \Gamma_{\mu ff'} i S_{f'} (\xi,0) \Gamma_{\nu ff'} q_f(0)
\right.
\nonumber \\
&& \qquad \qquad \left.+ \overline q_{f'}(0) \Gamma_{\nu ff'} iS_f(0,\xi) 
\Gamma_{\mu ff'} q_{f'}(\xi) \right\} |pS \rangle \quad.
\end{eqnarray} 
Here $S_f(\xi,0)$ is the fermionic propagator in the external 
field approximation for a quark with flavor $f$.
In the Schwinger formalism for calculations in external fields as proposed
by Shuryak and Vainshtein \cite{SV82}, (\ref{19}) can be written as
\begin{eqnarray}
\label{16}
T_{\mu \nu ff'} &=& - \langle pS|\left\{
      \overline q_f (0) \Gamma_{\mu ff'} \frac{1}{\Pslash - m_{f'} +\qslash}
         \Gamma_{\nu ff'} q_f(0) \right. 
\nonumber \\
&& \qquad \quad  \left. + \overline q_{f'}(0) \Gamma_{\nu ff'} 
          \frac{1}{\Pslash - m_{f} -\qslash}
         \Gamma_{\mu ff'} q_{f'}(0) \right\} |pS \rangle \quad,
\end{eqnarray}
with $P_\mu = i \partial_\mu +g A_\mu$.
Such an expression is very conveniently expanded in powers of $1/Q$
\begin{equation}
\label{17}
\frac{1}{\Pslash - m_f - \qslash} = - \frac{1}{\qslash}
         \sum_{j=0}^\infty \left( (\Pslash - m_f)\frac{1}{\qslash} \right)^j\quad.
\end{equation}
This is the starting point of our OPE analysis. 
%The l.h.s represents a
%non-local integral operator while on the r.h.s a sum over local
%differential operators appears. 
Inserting the expansion in (\ref{16}) it immediately 
becomes clear that for $f = f'$  all odd terms 
in $j$ are symmetric under exchange of $\mu$ and $\nu$
and thus contribute to the unpolarized structure functions, 
while
all even terms in $j$ are antisymmetric contributing to the 
polarized structure functions.

Finally, to define twist the different Lorentz structures 
are classified according to their
spin. Although this is standard in principle some ambiguities enter 
for more than two Lorentz indices. To avoid confusion we state explicitly the
decomposition we use in appendix A.

\section{Photon and $Z^0$ exchange} 
\noindent
In this section we perform the decomposition of the virtual forward Compton 
scattering amplitude for  $Z^0$ exchange.
The corresponding expression for $\gamma$ exchange 
can simply be obtained 
by setting $V_f \to Q_f$ and
$A_f \to 0$.
For parity-even quantities $\gamma$  and
$Z^0$ exchange always interfere. The interference term is given by
\begin{equation}
T_{\mu\nu}^{( \rm int\; \gamma, Z^0)}
= \left (T_{\mu\nu}^{(\gamma + Z^0)} \right)
        _{         V_f  \rightarrow  V_f + Q_f 
           \atop   A_f  \rightarrow  A_f }
- 
 \left (T_{\mu\nu}^{(\gamma)} \right)
       _{          V_f  \rightarrow  Q_f 
           \atop   A_f  \rightarrow  0 }
-           
 \left (T_{\mu\nu}^{(Z^0)} \right)
       _{          V_f  \rightarrow  V_f 
          \atop    A_f  \rightarrow  A_f } \quad.
\end{equation}
We will now proceed as in \cite{SV82} and \cite{E95}
except for using a different classification of matrix elements. 
The virtual forward Compton scattering amplitude
for $\gamma$ and $Z^0$ exchange reads
\begin{eqnarray}
\label{Eq30}
T^{(\gamma,Z^0)}_{\mu \nu} &=& 
 -\sum_f 
      \langle pS |\bigg\{ \overline q_f(0) \gamma_\mu
                   [V_f + A_f \gamma_5] \frac{1}{\Pslash -m_f + \qslash} \gamma_\nu
                   [V_f + A_f \gamma_5]q_f(0)
\nonumber \\
            &&  + \qquad \overline q_f(0) \gamma_\nu
                   [V_f + A_f \gamma_5]  \frac{1}{\Pslash -m_f - \qslash} \gamma_\mu
                   [V_f + A_f \gamma_5]q_f(\xi) \bigg\}|pS \rangle \quad.
\end{eqnarray}
Expanding in $(\Pslash- m_f)/Q$ according to (\ref{17}) we get for the 
zeroth order term 
\begin{eqnarray}
T^{(\gamma,Z^0)}_{\mu \nu} \bigg|_{\rm 0.\;{\rm order}}  &=& -\sum_f
      \langle pS | \bigg\{\overline q_f(0) \gamma_\mu
                   [V_f + A_f \gamma_5] \frac{\qslash}{q^2} \gamma_\nu
                   [V_f + A_f \gamma_5]q_f(0)
\nonumber \\
            &&  \qquad - \overline q_f(0) \gamma_\nu
                   [V_f + A_f \gamma_5]  \frac{\qslash}{q^2} \gamma_\mu
                   [V_f + A_f \gamma_5]q_f(0) \bigg\} |pS \rangle
\nonumber \\
&=&-2i \epsilon_{\mu \nu \lambda \sigma} \frac{q^\lambda S^\sigma}{\nu} \omega
     a_{f5}^{(0,2)} (V_f^2 + A_f^2)
\nonumber \\
&&-2i \epsilon_{\mu \nu \lambda \sigma} \frac{q^\lambda p^\sigma}{2\nu} \omega
     [2 a_{f}^{(0,2)}](2V_f A_f) \quad,
\end{eqnarray}
where the identity
\begin{equation}
\gamma_\mu \gamma_\alpha \gamma_\nu = g_{\mu \alpha}\gamma_\nu
+ g_{\nu \alpha} \gamma_\mu - g_{\mu \nu } \gamma_\alpha 
-i \epsilon_{\mu \alpha \nu  \sigma} \gamma_5 \gamma^\sigma 
\end{equation}
was used and the following matrix elements were defined:
\begin{eqnarray}
2a_{f5}^{(0,2)}S^\sigma &=& \langle pS |\overline q_f (0)
                   \gamma^\sigma \gamma_5
                                q_f (0) |pS \rangle \quad,
\nonumber \\
2a_{f}^{(0,2)}p^\sigma &=& \langle pS |\overline q_f (0)
                   \gamma^\sigma
                                q_f (0)|pS \rangle \quad.
\end{eqnarray}
The classification of matrix elements is to be understood as follows
\begin{equation}
\label{class}
a^{(m,t)}_{f5-} :\left\{ \begin{array}{cl}
              m = & {\rm number\;of\;}P_\mu-{\rm operators}. \\  
              t = & {\rm twist}.\\
              f = & {\rm quark\; flavor.}\\
              5 = & {\rm indicates \;\gamma_5 }.\\
              - = & {\rm indicates \;odd \;chirality,\; left\; out}\\
                  & {\rm in\; case\; of\; even\; chirality}.
          \end{array} \right. 
\end{equation}
The spin of the operator is 
$s = 3 + m - t$. The chirality of an operator is said to be even if
\begin{equation}
\overline q_f \gamma_5 \hat O \gamma_5 q_f  
= - \overline q_f  \hat O q_f 
\end{equation}
holds. The other case, with the positive sign, indicates chiral odd operators.
The contribution of first order is symmetric in $\mu$ and $\nu$. It reads:
\begin{eqnarray}
T^{(\gamma,Z^0)}_{\mu \nu} \bigg|_{\rm 1.\;{\rm order}}  &=& \sum_f
      \langle pS |\bigg\{ \overline q_f(0)
 \gamma_\mu [V_f + A_f \gamma_5] 
       \frac{\qslash}{q^2}(\Pslash - m_f)  \frac{\qslash}{q^2} 
 \gamma_\nu [V_f + A_f \gamma_5]q_f(0)
\nonumber \\
            &&  \qquad + \overline q_f(0) 
 \gamma_\nu [V_f + A_f \gamma_5]  
     \frac{\qslash}{q^2} (\Pslash - m_f)  \frac{\qslash}{q^2}
 \gamma_\mu [V_f + A_f \gamma_5]q_f(0)\bigg\} |pS \rangle \quad.
 \end{eqnarray}
We decompose the above expression into operators of well defined spin,
which are parametrized in terms of reduced matrix elements. 
We write the result in a way matching  the decomposition structure 
of the nucleon scattering tensor $W_{\mu \nu}$. In this way we get
\begin{eqnarray}
T^{(\gamma,Z^0)}_{\mu \nu} \bigg|_{\rm 1.\;{\rm order}}  
&=&  \left( - g_{\mu \nu} + \frac{q_\mu q_\nu}{q^2} \right) 
   \left[ \left( 2\omega^2 a_f^{(1,2)} 
         + 4\frac{M^2}{Q^2} a_f^{(1,2)}\right) (V_f^2 + A_f^2)
\right.
\nonumber \\
 && \qquad \qquad \qquad \qquad  \left.
         - 4\frac{m_f M}{Q^2} a_{f-}^{(0,3)}(V_f^2-A_f^2) \right]
\nonumber \\
&&+\frac{1}{\nu} \hat p_\mu \hat p_\nu 
   \left[ 4 \omega a_f^{(1,2)} (V_f^2 + A_f^2) \right]
+\frac{1}{\nu}q_\mu q_\nu 
  \left[  4 \omega \frac{m_f M}{Q^2}a_{f-}^{(0,3)} A_f^2  \right]  
\nonumber \\
&& +\left( - g_{\mu \nu} + \frac{q_\mu q_\nu}{q^2} \right) \frac{Sq}{\nu} 
  \left[ 2 \omega^2 a_{f5}^{(1,2)} \right]2V_f A_f
+q_\mu q_\nu \frac{Sq}{\nu^2} 
  \left[- \omega^3\frac{m_f}{M}  a_{f5-}^{(0,2)} \right]2V_f A_f
\nonumber \\
&&+\frac{Sq}{2\nu^2} (p_\mu q_\nu + p_\nu q_\mu) 
  \left[ - 2 \omega^2 \frac{m_f}{M}a_{f5-}^{(0,2)} \right]2V_f A_f
\nonumber \\
&&+\frac{1}{2\nu} ( \hat S_\mu \hat p_\nu +  \hat S_\nu \hat p_\mu )
  \left[4 \omega a_{f5}^{(1,2)} + 
        4 \omega  \frac{m_f}{M}a_{f5-}^{(0,2)} \right]2V_f A_f
\nonumber \\
&&+\frac{1}{2\nu} ( \hat S_\mu  p_\nu +  \hat S_\nu  p_\mu )
    \left[- 4\omega \frac{m_f}{M} a_{f5-}^{(0,2)} \right]2V_f A_f \quad.
\end{eqnarray}
For the matrix element $a_{f5-}^{(0,2)}$ we employed the equation of motion.
The matrix elements are again classified according to Eq.({\ref{class}). The
exact form of the operators can be found in 
appendix \ref{appA}.
\newline
The contribution of second order in $1/Q$ is 
antisymmetric in $\mu$ and $\nu$. It reads
\begin{eqnarray}
T^{(\gamma, Z^0)}_{\mu \nu} \bigg|_{\rm 2.\;{\rm order}}  &=& -\sum_f
      \langle pS | \overline q_f(0)
 \gamma_\mu [V_f + A_f \gamma_5]
       \frac{\qslash}{q^2}(\Pslash - m_f)  \frac{\qslash}{q^2}
                          (\Pslash - m_f)  \frac{\qslash}{q^2}
 \gamma_\nu [V_f + A_f \gamma_5]q_f(0)
\nonumber \\
            &&  \qquad - \overline q_f(0)
 \gamma_\nu [V_f + A_f \gamma_5]
     \frac{\qslash}{q^2} (\Pslash - m_f)  \frac{\qslash}{q^2}
                         (\Pslash - m_f)  \frac{\qslash}{q^2}
 \gamma_\mu [V_f + A_f \gamma_5]q_f(0) |pS \rangle
\nonumber \\
\end{eqnarray}
and can be decomposed into 
\begin{eqnarray}
\label{34}
T^{(\gamma, Z^0)}_{\mu \nu} \bigg|_{\rm 2.\;{\rm order}}
= - \frac{1}{q^6} \sum_f \bigg[ &&
  -8i\epsilon_{\mu \nu \lambda \sigma} q_\alpha q_\beta q^\lambda
    \langle pS| \overline q_f(0) P^\alpha P^\beta \gamma^\sigma q_f(0)
      \gamma_5 |pS \rangle (V_f^2 + A_f^2)
\nonumber \\ &&
+ 4q^2 q^\lambda \epsilon_{\mu \nu \lambda \sigma} 
                      \epsilon^\sigma_{\;\; \tau \rho \delta}   
    \langle pS| \overline q_f(0) P^\tau P^\rho \gamma^\delta q_f(0)   
                     |pS \rangle (V_f^2 + A_f^2)
\nonumber \\ &&
     + 2q^2 m_f q^\lambda \epsilon_{\mu \nu \lambda \sigma}
                         \epsilon^\sigma_{\;\; \tau \rho \delta}
 \langle pS| \overline q_f(0) \gamma^\tau \gamma^\rho P^\delta q_f(0)
                     |pS \rangle (V_f^2 + A_f^2)
\nonumber \\ &&
     + 4q^2 m_f q_\alpha   \langle pS| \overline q_f(0) P^\alpha 
      [\gamma_\mu \gamma_\nu - \gamma_\nu \gamma_\mu] q_f(0)
                     |pS \rangle A_f^2
\nonumber \\ &&
-4i \epsilon_{\mu \nu \lambda \sigma} q^2 q^\lambda m_f^2 
    \langle pS|\overline q_f(0) \gamma^\sigma \gamma_5 q_f(0) |pS \rangle
       (V_f^2 + A_f^2)
\nonumber \\ &&
-8i \epsilon_{\mu \nu \lambda \sigma} q_\alpha q_\beta q^\lambda 
    \langle pS| \overline q_f(0) P^\alpha P^\beta \gamma^\sigma q_f(0)
        |pS \rangle 2 A_f V_f
\nonumber \\ &&
+4 q^2 q^\lambda \epsilon_{\mu \nu \lambda \sigma}
                \epsilon^\sigma_{\;\; \alpha \beta \gamma} 
     \langle pS|\overline q_f(0) P^\alpha P^\beta \gamma^\gamma
     \gamma_5  |pS \rangle 2 A_f V_f \bigg] \quad.
\end{eqnarray}
Here we  
use the following
relations that result from the equation of motion $\Pslash q_f = m_f q_f$:
\begin{eqnarray}
\label{48}
m_f \langle pS| \overline q_f (0)\gamma_\mu q_f(0) |pS \rangle
&=& \langle pS| \overline q_f (0)P_\mu q_f(0) |pS \rangle \quad,
\nonumber \\
\langle pS| \overline q_f (0) \gamma_5 q_f(0) |pS \rangle &=& 0 \quad,
\nonumber \\
\langle pS| \overline q_f (0)P_\mu \gamma_5 q_f(0) |pS \rangle &=& 0 \quad,
\nonumber \\
\langle pS| \overline q_f (0)( P_\mu \gamma_\nu \gamma_\lambda 
+P_\nu \gamma_\lambda \gamma_\mu
+P_\lambda \gamma_\mu \gamma_\nu) \gamma_5 q_f (0) |pS \rangle
&=& 0 \quad.
\end{eqnarray}
The totally antisymmetric
spin-0 part has to be treated separately. 
It can be transformed in the following way (Note, that these equations
are only valid when sandwiched between 
$\langle pS|\overline q_f(0) $ and $ q_f(0) |pS \rangle$.):
\begin{eqnarray}
\label{equ1}
\epsilon^{\sigma \alpha \beta \gamma} P_\alpha P_\beta \gamma_\gamma 
&=& i (P^2 - m_f^2) \gamma^\sigma \gamma_5 \quad, 
 \\
\label{equ2}
\epsilon^{\sigma \alpha \beta \gamma} P_\alpha P_\beta \gamma_\gamma \gamma_5
&=& i (P^2 - m_f^2) \gamma^\sigma \quad,  
 \\
\label{equ3}
\epsilon^{\sigma \alpha \beta \gamma} \gamma_\alpha \gamma_\beta P_\gamma 
&=& 2i m_f \gamma^\sigma \gamma_5 \quad,
 \\
\label{equ4}
\epsilon^{\sigma \alpha \beta \gamma} \gamma_\alpha 
\gamma_\beta P_\gamma \gamma_5
&=& 0  \quad.                              
\end{eqnarray}
With the definition of the dual gluonic field strength tensor
$ig {\tilde G}_{\sigma \tau} = i\frac{g}{2} \epsilon_{\sigma \tau \alpha \beta} 
G^{\alpha \beta} =\frac{1}{2} \epsilon_{\sigma \tau \alpha \beta}
                [P_\alpha,P_\beta]$
and the equation of motion the operators $P^2 \gamma^\sigma$ and
$P^2 \gamma^\sigma \gamma_5$ can be transformed into a mass component and a 
gluonic component.
In terms of matrix elements this transformation reads:
\begin{eqnarray}
\label{gluon5}
m_f^2 a_{f5}^{(0,2)} + M^2 a_{f5}^{(2,4)} &=& M^2 \tilde a_{f5}^{(2,4)}\quad,
\\
\label{gluon}
m_f^2 a_{f}^{(0,2)} + M^2 a_{f}^{(2,4)} &=& M^2 \tilde a_{f}^{(2,4)} \quad,
\end{eqnarray} 
where the reduced matrix elements are listed in appendix \ref{appA}. Here also
a second relation is derived:
\begin{equation}
\label{anh22}
a_{f5}^{(2,3)} =  \tilde a_{f5}^{(2,3)}
-  \frac{m_f}{M} a_{f5-}^{(1,2)}
\quad.
\end{equation}
Collecting all terms we get for the second order of our expansion in 
$(\Pslash - m_f)/Q$:
\begin{eqnarray}
T^{(\gamma, Z^0)}_{\mu \nu} \Bigg|_{2.\;order}
 = -i \epsilon_{\mu \nu \lambda \sigma} 
                 \frac{q^\lambda S^\sigma}{\nu}
        && \left[ \frac{2}{3} \omega ^3 a_{f5}^{(2,2)}
                 +\frac{4}{3} \omega ^3 \tilde a_{f5}^{(2,3)}
                 +\frac{4}{9} \frac{M^2}{Q^2} \omega 
                \left(     a_{f5}^{(2,2)} 
                       + 4 \tilde a_{f5}^{(2,3)}
                       + 4 \tilde a_{f5}^{(2,4)} \right) \right. 
\nonumber \\
        && \left. - \frac{8}{3} \frac{1}{Q^2} \omega a_{f5}^{(0,2)} m_f^2
                     \right] (V_f^2 + A_f^2)
\nonumber \\
&& 
          -i \epsilon_{\mu \nu \lambda \sigma} 
             \frac{ q^\lambda S^\sigma}{\nu} 
             \left[ - \frac{4}{3} \frac{m_f^2}{Q^2} \omega a_{f5}^{(0,2)}
             \right] (V_f^2 - A_f^2) 
\nonumber \\
\nonumber \\
-i \epsilon_{\mu \nu \lambda \sigma}Sq
              \frac{q^\lambda p^\sigma}{\nu^2}  
       && \left[ \frac{4}{3} \omega^3 (a_{f5}^{(2,2)}- \tilde a_{f5}^{(2,3)})
                        (V_f^2 + A_f^2) \right]
\nonumber \\
\nonumber \\
-i \epsilon_{\mu \nu \lambda \sigma}\frac{q^\lambda p^\sigma}{2\nu}
&& \left[ 4 \omega^3 a_f^{(2,2)} + \frac{8}{9} \frac{M^2}{Q^2}
       \omega \left( 3a_f^{(2,2)} + 8 \tilde a_f^{(2,4)}\right) \right.
\nonumber \\
  && \left. -\frac{88}{9} \frac{1}{Q^2}\omega a_f^{(0,2)}
                 m_f^2 \right] 2 V_f A_f \quad.
\end{eqnarray} 
\section{$W^\pm$-exchange}
\noindent
In the case of $W^-$ exchange we use the current
\begin{equation}
j^{(W^-)}_\mu (\xi) = \overline d (\xi) \gamma_\mu (V - A \gamma_5) u(\xi)
+ \dots \quad,
\end{equation}
where the ellipsis denotes currents involving heavier quarks.
For simplicity we will discuss only the $ u \rightarrow d$ current.
For other flavors the relations look just the same.
%
%\begin{equation}
%j^{(W^+)}_\mu( \xi) = \overline u (\xi ) \gamma_\mu (V - A \gamma_5) d(\xi)
%+ \dots \quad. 
%\end{equation}
%
%
%For the following we concentrate on the $W^-$ exchange, 
%the case of $W^+$ exchange case is obtained simply by 
%exchanging $u$ and $d$ flavor.
%
%
The equation analogous to (\ref{Eq30}) is then given by
\begin{eqnarray}
\label{Tmunu}
T^{(W^-)}_{\mu \nu} &=& i \int d^4 \xi e^{iq\xi} \langle pS|T\bigg(
     \overline u (\xi) \gamma_\mu (V + A \gamma_5) d(\xi)
     \overline d (0) \gamma_\nu (V + A \gamma_5) u(0) \bigg)
|pS \rangle 
\nonumber \\
&=&
-\langle pS|\bigg\{ \overline u(0) \gamma_\mu (V + A \gamma_5)
            \frac{1}{ \Pslash - m_d + \qslash} \gamma_\nu (V + A \gamma_5) 
           u(0)
\nonumber \\
&& \qquad + \overline d(0) \gamma_\nu (V + A \gamma_5)
            \frac{1}{ \Pslash - m_u - \qslash} \gamma_\mu (V + A \gamma_5)
           d(0) \bigg\}|pS \rangle \quad,
\end{eqnarray}
where the expansion in $\Pslash - m_f$ 
becomes more complicated since two different flavors are involved. 
In each order symmetric and antisymmetric terms appear.
For the zeroth order term of the expansion we have
\begin{eqnarray}
\label{WM0order}
T^{(W^-)}_{\mu \nu} \bigg|_{0.\;{\rm  order}} &=&  
    - \langle pS | \{ 
       \overline u (0) \gamma_\mu (V+ A \gamma_5) 
                   \frac{\qslash}{q^2} \gamma_\nu (V+A \gamma_5) u(0)
\nonumber \\
&&  +     \overline d (0) \gamma_\nu (V+ A \gamma_5)
                   \frac{\qslash}{q^2} \gamma_\mu (V+A \gamma_5) d (0)
            \} |pS \rangle 
\nonumber \\
&=&
\left\{\left(-g_{\mu \nu} + \frac{q_\mu q_\nu}{q^2} \right)
\omega a_u^{(0,2)}
+ \frac{1}{2\nu} (p_\mu q_\nu + p_\nu q_\mu) [2\omega a_u^{(0,2)}]
+ \frac{1}{\nu} q_\mu q_\nu \frac{1}{2}\omega^2 a_u^{(0,2)}
\right. \nonumber \\ 
&& \left.-i\epsilon_{\mu \nu \lambda \sigma} 
\frac{q^\lambda S^\sigma}{\nu} \omega a_{u5}^{(0,2)} \right\}(V^2+ A^2)
\nonumber \\
&& +\left\{
\left(- g_{\mu \nu} + \frac{q_\mu q_\nu}{q^2} \right) 
\frac{Sq}{\nu}  \omega a_{u5}^{(0,2)} 
+ \frac{1}{2\nu} ( \hat S_\mu \hat p_\nu + \hat S_\nu \hat p_\mu)
[4 a_{u5}^{(0,2)}]
\right. \nonumber \\ 
&&-\frac{1}{2\nu} ( \hat S_\mu p_\nu + \hat S_\nu  p_\mu)
[4 a_{u5}^{(0,2)}]
+ \frac{Sq}{\nu^2} q_\mu q_\nu 
\left[- \frac{1}{2} \omega^2 a_{u5}^{(0,2)} \right]
\nonumber \\ 
&& \left. -i\epsilon_{\mu \nu \lambda \sigma}
\frac{q^\lambda p^\sigma}{2\nu} \omega [2a_{u}^{(0,2)}] \right\}2VA
+ (u \leftrightarrow d,\mu \leftrightarrow \nu, q \leftrightarrow -q )
\quad.
\end{eqnarray}
The fact that in first order of the expansion we get contributions
to current non-conserved structure functions, namely $F_4, F_5, b_2$, and $a_4$
is due to an incomplete definition of $T_{\mu \nu}$. Indeed, in zeroth
order no quark masses appear and consequently, both the
vector and the axial vector current should be conserved, requiring 
$ q^\mu T_{\mu \nu}|_{0.\;order} = 0$. The terms proportional to $F_4, F_5, b_2$, and $a_4$
are therefore unphysical and must be subtracted. 
In this sense our result (\ref{WM0order}) has to be rewritten in a form
where only current-conserved contributions occur:
\begin{eqnarray}
T^{(W^-)}_{\mu \nu} \bigg|_{0.\;{\rm  order}} &=&
\left\{\left(-g_{\mu \nu} + \frac{q_\mu q_\nu}{q^2} \right)
\omega a_u^{(0,2)}
\quad -i\epsilon_{\mu \nu \lambda \sigma}
\frac{q^\lambda S^\sigma}{\nu} \omega a_{u5}^{(0,2)} \right\}(V^2+ A^2)
\nonumber \\
&& +\left\{ 
\left(- g_{\mu \nu} + \frac{q_\mu q_\nu}{q^2} \right)
\frac{Sq}{\nu}  \omega a_{u5}^{(0,2)}
+ \frac{1}{2\nu} ( \hat S_\mu \hat p_\nu + \hat S_\nu \hat p_\mu)
[4 a_{u5}^{(0,2)}]
\right.
\nonumber \\
&& \left. -i\epsilon_{\mu \nu \lambda \sigma}
\frac{q^\lambda p^\sigma}{2\nu} \omega [2a_{u}^{(0,2)}] \right\}2VA
+ (u \leftrightarrow d,\mu \leftrightarrow \nu, q \leftrightarrow -q )
\quad.
\end{eqnarray}
This phenomenon arises only in the first order of expansion. Formally it
comes from unphysical contact terms
(seagull terms) generated by  derivatives acting on the theta function
in the time ordered product \cite{Bran70}.  
In this way we obtain, expanding the propagator to  first 
order 
\begin{eqnarray}
T^{(W^-)}_{\mu \nu} \bigg|_{1.\;{\rm  order}} &=&
     \langle pS | \{
       \overline u (0) \gamma_\mu (V+ A \gamma_5)
       \frac{\qslash}{q^2} (\Pslash - m_d) \frac{\qslash}{q^2} 
       \gamma_\nu (V+A \gamma_5) u(0)
\nonumber \\
&& +   \overline d (0) \gamma_\nu (V+ A \gamma_5)
       \frac{\qslash}{q^2} (\Pslash - m_d) \frac{\qslash}{q^2} 
       \gamma_\mu (V+A \gamma_5) d (0)
            \} |pS \rangle
\nonumber \\
& = &
\left\{
  \left( -g_{\mu \nu} + \frac{q_\mu q_\nu}{q^2} \right)
  \left[ \omega^2 a_u^{(1,2)} + 2 \frac{M^2}{Q^2} a_u^{(1,2)} \right]
\right.
\nonumber \\
&& \left.
+  \frac{1}{\nu} \hat p_\mu \hat p_\nu [ 2 \omega a_u^{(1,2)} ]
+  \frac{1}{\nu} q_\mu q_\nu \left[ \frac{m_u M}{Q^2} \omega a_{u-}^{(0,3)} 
   \right] \right\} (V^2 + A^2)
\nonumber \\
&& + \left\{
\left(-g_{\mu \nu} + \frac{q_\mu q_\nu}{q^2} \right)\frac{Sq}{\nu}
[\omega^2 a_{u5}^{(1,2)}]
+
\frac{Sq}{2\nu^2} (p_\mu q_\nu + p_\nu q_\mu) 
[-\omega^2 \frac{m_u}{M}a_{u5-}^{(0,2)}] \right.
\nonumber \\
&& + \frac{Sq}{\nu^2}q_\mu q_\nu  
    \left[-\frac{1}{2} \omega^3 \frac{m_u}{M}a_{u5-}^{(0,2)} \right]
 + \frac{1}{2\nu} ( \hat S_\mu \hat p_\nu + \hat S_\nu \hat p_\mu)
   [2 \omega (a_{u5}^{(1,2)} + \frac{m_u}{M}a_{u5-}^{(0,2)} ) ]
\nonumber \\
&& \left.
+ \frac{1}{2\nu} ( \hat S_\mu  p_\nu + \hat S_\nu  p_\mu)
   [-2 \omega \frac{m_u}{M}a_{u5-}^{(0,2)} ]
 \right\}2VA  
\nonumber \\
&&+\left(- i \epsilon_{\mu \nu \lambda \sigma }
\frac{q^\lambda S^\sigma}{\nu} 
\left[
     \frac{\omega^2}{2} ( a_{u5}^{(1,2)}  + \frac{m_u}{M}a_{u5-}^{(0,2)}) 
\right] \right.
\nonumber \\
&&\quad - i \epsilon_{\mu \nu \lambda \sigma }
Sq \frac{q^\lambda p^\sigma}{\nu^2}
\left[
\frac{\omega^2}{2} ( a_{u5}^{(1,2)} - \frac{m_u}{M}a_{u5-}^{(0,2)}) 
\right] 
\nonumber \\
&& \quad  - i \epsilon_{\mu \nu \lambda \sigma } 
\frac{p^\lambda S^\sigma}{\nu} [\omega \frac{m_u}{M}a_{u5-}^{(0,2)}] \Bigg) (V^2 + A^2)
\nonumber \\
&&\quad -i \epsilon_{\mu \nu \lambda \sigma }
\frac{q^\lambda p^\sigma}{2\nu} [2\omega^2 a_{u}^{(1,2)}] 2VA       
\nonumber \\
&&+ (u \leftrightarrow d , \mu \leftrightarrow \nu) \quad.
\end{eqnarray}
The expression for the second order term 
\begin{eqnarray}
T^{(W^-)}_{\mu \nu} \bigg|_{{2.\;{\rm  order}}} &=&
     \langle pS | \{
       \overline u (0) \gamma_\mu (V+ A \gamma_5)
       \frac{\qslash}{q^2} (\Pslash - m_d)
       \frac{\qslash}{q^2} (\Pslash - m_d) \frac{\qslash}{q^2}
       \gamma_\nu (V+A \gamma_5) u(0)
\nonumber \\
&& +   \overline d (0) \gamma_\nu (V+ A \gamma_5)
       \frac{\qslash}{q^2} (\Pslash - m_u)  
       \frac{\qslash}{q^2} (\Pslash - m_u) \frac{\qslash}{q^2}
       \gamma_\mu (V+A \gamma_5) d (0)
            \} |pS \rangle
\nonumber
\end{eqnarray}
is again decomposed into components antisymmetric and symmetric 
in $\mu$ and $\nu$. The latter one reads:
\begin{eqnarray}
&T^{(W^-)}_{\mu \nu} \bigg|_{{2.\;{\rm  order}}
\atop  {\rm sym.\;in \;\mu \nu }} =
\Bigg\{
\left(-g_{\mu \nu} + \frac{q_\mu q_\nu}{q^2} \right)
\left[
\omega^3 a_u^{(2,2)} + \frac{8}{9} \frac{M^2}{Q^2}\omega 
                        [3a_u^{(2,2)} - \tilde a_u^{(2,4)}]
\right. \nonumber \\ & \left.
      + \frac{1}{9} \frac{1}{Q^2} [11 m_u^2 -9 m_d^2] \omega a_u^{(0,2)}
\right]
+ \frac{1}{2\nu} (p_\mu q_\nu + p_\nu q_\mu) 
\left[ \frac{2}{Q^2} [m_u^2 - m_d^2]\omega a_u^{(0,2)} \right]
\nonumber \\
&+ \frac{1}{\nu}q_\mu q_\nu
     \left[
      \frac{1}{2}  \frac{\omega^2}{Q^2}[3m_u^2 - m_d^2] a_u^{(0,2)} 
      \right]
+\frac{1}{\nu} \hat p_\mu \hat p_\nu 
\left[ 2 \omega^2 a_u^{(2,2)} \right] \Bigg\}
 (V^2 + A^2)
\nonumber \\
\nonumber \\
&+ \left\{\left( -g_{\mu \nu} + \frac{q_\mu q_\nu}{q^2} \right)
\frac{Sq}{\nu} \left[ \omega^3 a_{u5}^{(2,2)} 
                      +\frac{8}{9} \frac{M^2}{Q^2} \omega 
                [ a_{u5}^{(2,2)} - 2 \tilde a_{u5}^{(2,3)} + \tilde a_{u5}^{(2,4)}]
\right.\right. \nonumber \\&  \qquad \qquad\left.
+\frac{4}{3} \frac{m_u M}{Q^2} \omega a_{u5-}^{(1,2)}
-\frac{1}{3} \frac{1}{Q^2}  [m_u^2 +3 m_d^2]\omega  a_{u5}^{(0,2)} \right]
\nonumber \\
&+ \frac{Sq}{2\nu^2}(p_\mu q_\nu + p_\nu q_\mu)
\left[- \frac{m_u}{M} \omega^3 a_{u5-}^{(1,2)} \right]
+ \frac{Sq}{\nu^2} q_\mu q_\nu 
\left[
 - \frac{1}{2} \frac{m_u}{M} \omega^4 a_{u5-}^{(1,2)}
 - \frac{1}{2} \frac{1}{Q^2} [m_u^2 - m_d^2] \omega^2 a_{u5}^{(0,2)}
\right]
\nonumber \\
\nonumber \\
&+\frac{Sq}{\nu^2} \hat p_\nu \hat p_\nu 
    \left[ \frac{2}{3} \omega^2 a_{u5}^{(2,2)}
          -\frac{8}{3} \omega^2 \tilde a_{u5}^{(2,3)}
          + 2\frac{m_u}{M} \omega^2 a_{u5-}^{(1,2)} \right]
\nonumber \\
&+ \frac{1}{2\nu} ( \hat S_\mu \hat p_\nu + \hat S_\nu \hat p_\mu)
\left[  \frac{4}{3} \omega^2 a_{u5}^{(2,2)}
       +\frac{8}{3} \omega^2 \tilde a_{u5}^{(2,3)}
+ \frac{8}{3} \frac{m_u M}{Q^2} a_{u5-}^{(1,2)}
+ \frac{4}{3} \frac{1}{Q^2} [m_u^2 - 3 m_d^2] a_{u5}^{(0,2)}
\right]
\nonumber \\
&  
+\frac{1}{2\nu} ( \hat S_\mu  p_\nu + \hat S_\nu  p_\mu) \left[
- 2 \frac{m_u}{M} \omega^2 a_{u5-}^{(1,2)} 
- \frac{8}{3} \frac{m_u M}{Q^2} a_{u5-}^{(1,2)} \left.
- \frac{4}{3} \frac{1}{Q^2} [ m_u^2 - 3 m_d^2] a_{u5}^{(0,2)}
\right]  \right\} 2VA
\nonumber \\
&+ (u \leftrightarrow d , \mu \leftrightarrow \nu, q \leftrightarrow -q )
\quad.
\end{eqnarray}   
The corresponding antisymmetric part differs from the $Z^0$ case only
by the mass terms and, of course, by a factor 1/2.
\begin{eqnarray}
T^{(W^-)}_{\mu \nu} \bigg|_{{2.\;{\rm  order}} \atop 
{ \rm asymm.\; in \;\mu \nu}} &=&
\left\{-i \epsilon_{\mu \nu \lambda \sigma}
\frac{q^\lambda S^\sigma}{\nu}
\left[ \frac{1}{3} \omega^3 a_{u5}^{(2,2)} 
     + \frac{2}{3} \omega^3 \tilde a_{u5}^{(2,3)} \right. \right.
\nonumber \\ &&\left. 
\qquad \quad+ \frac{2}{9} \frac{M^2}{Q^2} \omega
          [a_{u5}^{(2,2)} + 4 \tilde a_{u5}^{(2,3)} + 4 \tilde a_{u5}^{(2,4)}]
      - \frac{1}{3} \frac{1}{Q^2}[m_u^2 +3 m_d^2] a_{u5}^{(0,2)} \omega
\right]
\nonumber \\
\nonumber \\&& \left.
-i \epsilon_{\mu \nu \lambda \sigma}
\frac{q^\lambda p^\sigma}{\nu^2}Sq 
\left[ \frac{2}{3} \omega^3[ a_{u5}^{(2,2)} - \tilde a_{u5}^{(2,3)}] \right]
\right\}(V^2 + A^2)
\nonumber \\
\nonumber \\&&
+ \left\{-i \epsilon_{\mu \nu \lambda \sigma}
\frac{q^\lambda p^\sigma}{2\nu} 
\left[ 2 \omega^3 a_u^{(2,2)} 
      + \frac{4}{9}\frac{M^2}{Q^2}\omega [3a_u^{(2,2)}+ 8\tilde a_u^{(2,4)}]
\right. \right. \nonumber \\&& \qquad \qquad \left. \left.
      - \frac{1}{9}\frac{1}{Q^2} [26 m_u^2 + 18 m_d^2] a_u^{(0,2)} \omega
\right] \right\}2VA
\nonumber \\
\nonumber \\
&&+ (u \leftrightarrow d , \mu \leftrightarrow \nu, q \leftrightarrow -q )
\quad.
\end{eqnarray}
We have discussed matrix elements up to twist 4. At this twist
level the cat ear diagram (see figure 4) may contribute. It describes
non-perturbative 
interactions between the hit quark and the spectator quarks in the final state.
However, it turns out (see appendix B) 
that the matrix elements of the cat ear diagram
yield contributions of structure functions that are one order 
in $1/Q^2$ higher such that we do not have to include them in the order we are
working. Also, due to symmetry properties cat ear diagrams can only contribute 
to $F_1$, $F_2$, $a_1$, and $b_2$, as discussed in appendix B. 
\section{Sum rules}
\noindent
Having analyzed  the virtual forward Compton scattering amplitude  
completely up to order $[(\Pslash - m_f)/Q]^2$, 
we have determined the lowest moments of the 14 structure functions.
The results are compiled in table II and the resulting sum rules are listed in 
table III on the leading twist level. The mass corrections to these sum rules can 
be read off from table II so we do not repeat them. Table IV gives some relations 
between structure functions, valid in the limit $Q^2\rightarrow \infty$, i.e.
neglecting higher-twist corrections. In general, here we do not consider 
possible anomalous contributions arising from the divergence of pseudovector
currents. \\

Let us add some comments to these tables:\\

1.) The index $f$ runs over all quark and
antiquark flavors, i.e. $f = u,\overline u,d,\overline d, \dots$
Due to the opposite helicity of quarks and
antiquarks we have
\begin{equation}
V_f = V_{\overline f} \quad ; \qquad A_f = - A_{\overline f} \quad.
\end{equation}
Furthermore, we use in the case of neutrino scattering
for the weak coupling constants the abbreviation
\begin{equation}
\eta_W = \frac{1}{8 \sin^2 \theta_W} = \frac{G}{\sqrt{2}} 
\frac{ M_W^2}{4 \pi \alpha} \quad,
\end{equation}
and for brevity, in the case of $W^\pm$ - exchange, we introduce the 
following flavor combinations: $S$ (singlet), $V$ (valence) and furthermore
the combinations $\Delta S$ and $\Delta V$.
\begin{eqnarray}
a_S &:=& \eta_W[ (a_u + a_{\overline u}) + (a_d + a_{\overline d}) 
+ (\rm further\; generations)]\;,
\nonumber \\
a_V &:=& \eta_W[ (a_u - a_{\overline u}) + (a_d - a_{\overline d})
+ (\rm further\; generations)]\;,
\nonumber \\
a_{\Delta S} &:=& 
         \eta_W[ (a_u + a_{\overline u}) - (a_d + a_{\overline d})      
+ (\rm further\; generations)]\;,
\nonumber \\
a_{\Delta V}
     &:=& \eta_W[ (a_u - a_{\overline u}) - (a_d - a_{\overline d})      
+ (\rm further\; generations)]\;.
\end{eqnarray}
The relations between these flavor combinations and the 
symmetries of the amplitudes are presented in table \ref{tabflav}. 

To simplify the mass terms we define 
\begin{eqnarray}
  \left( m_S - {m'}_S \right)a_S 
&=&   \eta_W [\left( m_u - m_d \right)a_u 
         + \left( m_{\overline u} - m_{\overline d} \right)a_{\overline u}
\nonumber \\
&& +    \left( m_d - m_u \right) a_d
         + \left( m_{\overline d} - m_{\overline u} \right)a_{\overline d} 
+ (\rm further\; generations)]\;,
\end{eqnarray} 
and analogous expressions for V, $\Delta$ S, and $\Delta$ V. 

The constant $C$ appearing in table III for the 
polarized Bjorken sum rule has the value
\begin{equation}
\label{86}
C = \underbrace{\frac{1}{3}}_{\gamma}
   +\underbrace{\frac{1}{3 \sin 2 \theta_W}[4  \sin \theta_W -1]}_
               { \gamma - Z^0 \; \rm interference}
   +\underbrace{\frac{2}{3} \frac{\sin \theta_W}{\sin^2 2 \theta_W}
                  [2 \sin \theta_W -1]}_
               { Z^0}
 \quad.   
\end{equation}

Isospin  $T_3$ and baryon number $B$ are given by
\begin{eqnarray}
\label{EQ75}
\sum_{f= u,c,t}  \left(  a_f^{(0,2)}- a_{\overline f}^{(0,2)} \right)
-\sum_{f = d,s,b}  \left(  a_f^{(0,2)}- a_{\overline f}^{(0,2)} \right) = 2 T_3
\quad,
\\
\label{EQ75a}
\sum_{f= u,c,t}  \left(  a_u^{(0,2)}- a_{\overline f}^{(0,2)} \right)
+\sum_{f = d,s,b}  \left(  a_f^{(0,2)}- a_{\overline f}^{(0,2)} \right) = 3B
\quad,
\end{eqnarray}

Finally we note that we do not include here the radiative 
corrections and chose for simplicity the normalization point
to be $\mu^2= Q^2$. \\

2.) The twist-2 
matrix element of the tensor operator $(-i) \sigma_{\mu \nu}$ 
determines the transversity 
distribution $h_1$ \cite{RJa91a} and does not appear at leading twist 
in deep inelastic lepton nucleon scattering. As a chiral-odd operator 
it is related to spin-flip processes which are always suppressed by the
quark mass. As may be expected beforehand this operator determines the
polarized structure function $g_3$ which is related to a non-conserved 
current and vanishes identically for massless quarks.\\

3.) The simplest scalar twist-3 operator
is the scalar operator 
\beq{e4}
\langle p S | \overline q_f (0)  q_f (0)| p S \rangle = 2 a_{f-}^{(0,3)}M
\quad,
\eeq
which for the flavor combination 
$\frac{1}{2} (m_u + m_d) (\bar u u + \bar d d)$ is known as the
$\sigma$-term, giving rise to a sum rule for $F_4$ which, however is expected 
to be divergent due to a Regge analysis \cite{RJa91a}.
Since the scalar operator is chiral odd it will  appear in deep inelastic
scattering only if quark masses are not ignored.

4.) The twist-3  quark-gluon-quark correlator
\begin{eqnarray}
\label{qgq}   
&&\langle p S | \overline q_f (0) \frac{1}{6}
 \Bigg[ (\gamma_\alpha g \tilde G_{\beta \sigma} +
\gamma_\beta  g \tilde G_{\alpha \sigma}) \Bigg]
q_f (0) | p S \rangle   - {\rm traces}
\nonumber \\
&=& 2     a_{f5}^{(2,3)} \Bigg[\frac{1}{3}
                            (2   p_\alpha p_\beta  S_\sigma
                               - p_\beta  p_\sigma S_\alpha
                               - p_\sigma p_\alpha S_\beta)
- {\rm traces} \Bigg]
\end{eqnarray}
(often referred to
as $d^{(2)}$) contributes to the
first moment of $g_1$ and the third moment of
$g_2$. This operator contributes also higher-twist 
terms to $a_1$ and $a_2$.\\

5.) The twist-4 quark-gluon-quark correlators
\beq{qgq2}
\langle p S | \overline q_f (0) g \tilde G^{\alpha \beta}
\gamma_\beta q_f (0) | p S \rangle 
= 2  a_{f5}^{(2,4)} M^2 S^\alpha \quad.
\eeq
contributes to $g_1$ and $a_1$ and
this matrix element is commonly referred to as $f^{(2)}$.
The spin independent twist-4 operator  defined as
\beq{qgq3}
\langle p S | \overline q_f (0) g \tilde G^{\alpha \beta}
\gamma_\beta \gamma_5 q_f (0) | p S \rangle
= 2  a_{f}^{(2,4)} M^2 p^\alpha \quad,
\eeq
gives a higher-twist corrections to the Gross-Llewellyn Smith
sum rule.
(The matrix element of the analogous twist-3
operator vanishes due to its symmetry properties, i.e. $a_f^{(2,3)} 
\equiv 0$.)\\

6.) It is possible to relate the higher-twist contributions
for the second moment of $g_2$ to the structure function $g_3$
and to obtain an interesting relation, which reads
\begin{equation}
\int_0^1 x dx g_2^{(\overline \nu - \nu)} = - \frac{1}{2}
\int_0^1 x dx g_1^{(\overline \nu - \nu)}
+ \frac{1}{2} \int_0^1 dx g_3^{(\overline \nu - \nu)} +
{\cal O} (1/Q^2)
\quad.
\end{equation}

7.) We cannot confirm the relation
$ a_2(x) = 2x a_1(x)$ for  $Q^2 \to \infty$,
which was claimed in \cite{Ji93}, but rather reproduce the relation found
by Ravishankar \cite{Ra92} and earlier by Dicus \cite{Di72}:
\begin{equation}
2x a_1(x) =  a_2(x) +  b_1(x) +  b_2(x) \quad.
\end{equation}

8.) For the polarized parity-non-conserving structure functions we 
reproduce the relation derived by Wray \cite{Wr72} by means of the
light-cone quark algebra:
\begin{eqnarray}
\int_0^1 dx 
\left( a_1 ^{(\overline \nu - \nu)p}
      -a_1 ^{(\overline \nu - \nu)n} \right) 
&=& - 2  \frac{g_A}{g_V}\eta_w + {\cal O} (1/Q^2) \quad.
\end{eqnarray}

While the current non-conserved
structure functions, i.e. $a_4,a_5$, and $b_2$, will hardly be measurable
in practical experiments, the current conserved structure functions 
$a_1,a_2$ and $b_1$ should be experimentally accessible \cite{An94}.
They give interesting additional information on the
structure of the nucleon.

9.) We would like to stress that the operators 
connected with $ a_{f5}^{(2,3)} $ and $a_{f5}^{(2,4)}$
measure important
properties of the nucleon, namely contributions of the collective 
color electromagnetic field to the spin and to the momentum of the nucleon.
In the rest system $S^\alpha = (0,\vec{S})$, $p^\alpha = (M,\vec{0})$ the 
spin-dependent operators can be written as \cite{Stein2}
\begin{eqnarray}
\langle pS|  \vec j_a \cdot \vec B  |pS \rangle &=& 0 \quad,
\nonumber \\
\langle pS|\left[
- \vec B_a j_a^0 + (\vec j_a \times \vec E_a )
\right]
  |pS \rangle 
&=& 2 M^2 a_{f5}^{(2,4)} \vec S \quad,
\nonumber \\
\langle pS|
\left[
 2 \vec B_a j_a^0 + (\vec j_a \times \vec E_a ) \right] |pS \rangle 
&=& 8 M^2 a_{f5}^{(2,3)} \vec S \quad,
\end{eqnarray}
where $j^\mu_a = -g \overline q \gamma^\mu t_a q$ denotes the quark
current, $B^\sigma_a$ the color magnetic field,
and $E_a^\sigma$ the color electric field.
$ a_{f5}^{(2,3)} $ and $a_{f5}^{(2,4)}$ were calculated in relativistic
quark models such as the MIT bag model \cite{JiU94} as well as in the framework
of QCD sum rules \cite{BBK,Stein1,Stein2} making it possible to 
estimate the strength of the color  
electric and magnetic fields within the nucleon separately.

Analogously,  for the axial vector 
current $j^\mu_{a5} = -g \overline q \gamma^\mu \gamma_5t_a q$ we get 
\begin{eqnarray}
\langle pS|
 \vec j_{a5} \cdot \vec B |pS \rangle  &=& - 2M^3 a_{f}^{(2,4)} \quad, 
\nonumber \\  
\langle pS| \left[
 - \vec B_a j_{a5}^0 + (\vec j_{a5} \times \vec E_a ) 
\right]
|pS \rangle 
&=& 0 \quad,
\nonumber \\
\langle pS|  \left[
2 \vec B_a j_{a5}^0 + (\vec j_{a5} \times \vec E_a ) 
 \right] |pS \rangle 
&=& 0\quad.        
\end{eqnarray}
The two operators  $\vec j_{a5} \times \vec E_a$ and 
$\vec B_a j_{a5}^0$ determine a vector direction. As in 
the nucleon  rest frame
the only distinct direction is given by the spin, which is an axial vector,
the matrix elements of $\vec j_{a5} \times \vec E_a$ and $\vec B_a j_{a5}^0$
must vanish. We would like to point out, that the only nonvanishing matrix
element in this context, namely $a_f^{(2,4)}$, determines 
the higher-twist corrections to the Gross-Llewellyn Smith sum rule and
was calculated in the QCD sum rule approach in \cite{Bra87}.  

\section{summary and conclusions}
\noindent
In this paper we have presented a complete analysis for
the first moments of the structure functions
of the nucleon scattering tensor for $\gamma$, $Z^0$,
and $W^\pm$ exchange. 
All our calculations include the effects
of finite quark masses.\\
Besides  reproducing  well-known results
we found the relation
\begin{equation}
b_2(x) = 2x a_5(x) = 4x^2 a_4(x) \quad.
\end{equation}
The extension of our work to the third order in $\Pslash - m_f$
will require a consistent treatment of the cat ear diagrams.
\newline
\newline
{\bf Acknowledgements}
\newline
We thank Lech Mankiewicz for many
useful discussions and continuous encouragement.  This work has been  
supported by DFG (G.~Hess Programm), Cusanuswerk, and BMBF.  A.S. thanks 
the MPI f\"ur Kernphysik in Heidelberg for support.
\begin{appendix}
\section{Definition of the matrix elements}
\label{appA}
\noindent
For the sake of brevity we introduce the notation of the totally 
symmetric tensor $
s_{\alpha_1 \alpha_2 \alpha_3 \alpha_4} := 
|\epsilon_{\alpha_1 \alpha_2 \alpha_3 \alpha_4}|$, i.e. $s_{\alpha_1 \alpha_2 \alpha_3 \alpha_4}$
vanishes for the same index combinations as 
$\epsilon_{\alpha_1 \alpha_2 \alpha_3 \alpha_4}$, 
and whenever there is an index combination such that 
$\epsilon_{\alpha_1 \alpha_2 \alpha_3 \alpha_4} = \pm 1$,
then $s_{\alpha_1 \alpha_2 \alpha_3 \alpha_4} = 1$.
All other conventions are taken from \cite{itz80}.
\begin{itemize}
\item[i)] $g^{\mu \nu}$ has spin 0.
\item[ii)] Each Lorentz vector $P^\mu$ and $\gamma^\mu$ has spin 1.
\item[iii)] 
The operator $P^\mu \gamma^\nu$
can be decomposed into a symmetric and traceless spin-2 part, 
an antisymmetric spin-1 part,  and a spin-0 part according to
\begin{eqnarray}
P^\mu \gamma^\nu &=& 
\left[P^\mu \gamma^\nu\right]_{\rm Spin-2} + 
\left[P^\mu \gamma^\nu\right]_{\rm Spin-1} + 
\left[P^\mu \gamma^\nu\right]_{\rm Spin-0}
\nonumber \\
&=& 
\left[
\frac{1}{2}  \left(P^\mu \gamma^\nu + P^\nu \gamma^\mu \right)
- \frac{1}{4} g^{\mu \nu} \Pslash \right]
+ 
\left[
\frac{1}{2}  \left(P^\mu \gamma^\nu - P^\nu \gamma^\mu \right)
\right]
+
\left[
\frac{1}{4} g^{\mu \nu} \Pslash\right]
\quad.
\end{eqnarray} 
\item[iv)] The operator $P^\alpha P^\beta \gamma^\gamma$
is  decomposed into spin-3, spin-2, spin-1 and spin-0 parts.
There is one ambiguity in the definition of the spin-0 part:
Sandwiched between nucleon states, it can be transformed into a
spin-1 operator via the equation of motion
(see Eqns. (\ref{equ1}-\ref{equ3})). 
\begin{eqnarray}
\label{36}
P^\alpha P^\beta \gamma^\gamma &=& 
\left[P^\alpha P^\beta \gamma^\gamma \right]_{\rm Spin-3} +
\left[P^\alpha P^\beta \gamma^\gamma \right]_{\rm Spin-2} +
\left[P^\alpha P^\beta \gamma^\gamma \right]_{\rm Spin-1} +
\left[P^\alpha P^\beta \gamma^\gamma \right]_{\rm Spin-0} 
\nonumber \\
&=& 
\Bigg[\frac{1}{6} 
 s^{\alpha \beta \gamma \delta} s_{\delta \sigma \tau \lambda}
 P^\sigma P^\tau \gamma^\lambda
\nonumber \\
&& -  \frac{1}{18} \left(
         g^{\alpha \beta}[ P^2 \gamma^\gamma 
                   + \Pslash P^\gamma + P^\gamma \Pslash ] \right.
                   +  g^{\alpha \gamma}[ P^2 \gamma^\beta
                   + \Pslash P^\beta + P^\beta \Pslash ]
\nonumber \\
&& \qquad \quad \left. 
       +  g^{\gamma \beta}[ P^2 \gamma^\alpha 
                   + \Pslash P^\alpha + P^\alpha \Pslash ]         
\right)
\Bigg]
\nonumber \\
&&+ \Bigg[
\frac{1}{3} \left(   2 P^\alpha P^\beta   \gamma^\gamma
                       -   P^\gamma P^\alpha  \gamma^\beta 
                       -   P^\beta  P^\gamma   \gamma^\alpha  \right)
\nonumber \\
&& - \frac{1}{9} \left(
         g^{\alpha \beta}[2 P^2 \gamma^\gamma
                   - \Pslash P^\gamma - P^\gamma \Pslash] \right.
    +  g^{\alpha \gamma}[ 2 \Pslash P^\beta
                   - P^\beta \Pslash  - P^2 \gamma^\beta]
\nonumber \\
&& \qquad \quad \left.
      +  g^{\gamma \beta}[ 2 P^\alpha \Pslash
                   - \Pslash P^\alpha - P^2 \gamma^\alpha] \right) 
\Bigg]
\nonumber \\
&& + 
\Bigg[
\frac{1}{18} \left( 
         g^{\alpha \beta}[5P^2 \gamma^\gamma
                   - \Pslash P^\gamma - P^\gamma \Pslash] \right.
      +  g^{\alpha \gamma}[5 \Pslash P^\beta
                   - P^\beta \Pslash - P^2 \gamma^\beta]
\nonumber \\
&& \qquad \quad \left.
      +  g^{\gamma \beta} [5 P^\alpha \Pslash 
                   -\Pslash P^\alpha - P^2 \gamma^\alpha] \right)
\Bigg]
\nonumber \\
&& + \Bigg[
\frac{1}{6}
 \epsilon^{\alpha \beta \gamma \delta} \epsilon_{\delta \sigma \tau \lambda}
 P^\sigma P^\tau \gamma^\lambda \quad 
\Bigg] \quad .
\end{eqnarray}  
\item[v)] For the operator $\gamma^\alpha \gamma^\beta P^\gamma$
the situation is essentially the same as in iv), however,
this operator has no spin-3 part due to the fact that different gamma
matrices anticommute.
\begin{eqnarray}
\gamma^\alpha \gamma^\beta P^\gamma &=&
\left[\gamma^\alpha \gamma^\beta P^\gamma\right]_{\rm Spin-2} + 
\left[\gamma^\alpha \gamma^\beta P^\gamma\right]_{\rm Spin-1} + 
\left[\gamma^\alpha \gamma^\beta P^\gamma\right]_{\rm Spin-0}
\nonumber \\
& =& \Bigg[\frac{1}{6} 
 \left(   2 \gamma^\alpha \gamma^\beta  P^\gamma
       -    \gamma^\beta  \gamma^\gamma P^\alpha
       -    \gamma^\gamma \gamma^\alpha P^\beta  
       -  2 \gamma^\beta  \gamma^\alpha  P^\gamma
       +    \gamma^\gamma  \gamma^\beta P^\alpha
       +    \gamma^\alpha  \gamma^\gamma P^\beta  \right)
\nonumber \\
&&
- \frac{1}{6} \left( 
   g^{\alpha \gamma} 
        [\Pslash \gamma^\beta - \gamma^\beta \Pslash]
  -g^{\beta \gamma}
        [\Pslash \gamma^\alpha - \gamma^\alpha \Pslash] \right)
\Bigg]
\nonumber \\
&&+ \Bigg[\frac{1}{6} \left(
   g^{\alpha \gamma}  
        [\Pslash \gamma^\beta - \gamma^\beta \Pslash]
  -g^{\beta \gamma}
        [\Pslash \gamma^\alpha - \gamma^\alpha \Pslash] \right)
  + g^{\alpha \beta} P^\gamma \Bigg]
\nonumber \\
&& + \Bigg[
\frac{1}{6}
 \epsilon^{\alpha \beta \gamma \delta} \epsilon_{\delta \sigma \tau \lambda}
 \gamma^\sigma \gamma^\tau P^\lambda \Bigg] \quad .
\end{eqnarray}
\end{itemize}

According to this decomposition we define the various matrix elements
in the following manner, ordered according to the number of vector indices.\\

\noindent
\underline{Scalar and pseudoscalar operators}
\newline
\begin{eqnarray}
\label{anh1}
\langle p S | \overline q_f (0)  q_f (0) | p S \rangle
&=& 2 a_{f-}^{(0,3)} M \quad, \\
\label{anh2}
\langle p S | \overline q_f (0) \gamma_5 q_f (0) | p S \rangle &=&  0 
\quad.
\end{eqnarray}
\newline 
\underline{Vector and pseudovector operators} 
\begin{eqnarray}
\label{anh3}
\langle p S | \overline q_f (0)  P_\mu q_f (0) | p S \rangle 
&=&  m_f \langle p S | \overline q_f (0)  \gamma_\mu q_f (0) | p S \rangle
\quad, \\
\label{anh4}
\langle p S | \overline q_f (0)  P_\mu \gamma_5 q_f (0) | p S \rangle
&=& 0 \quad.
\end{eqnarray}
The last equation was simplified using the equation of motion. 
\begin{eqnarray}
\label{anh5}
\langle p S | \overline q_f (0)  \gamma_\mu q_f (0) | p S \rangle
&=& 2 a_f^{(0,2)} p_\mu \quad, \\
\label{anh6}
\langle p S | \overline q_f (0)  \gamma_\mu \gamma_5 q_f (0) | p S \rangle
&=& 2 a_{f5}^{(0,2)} S_\mu \quad.
\end{eqnarray}
\newline
\underline{Rank-two tensor operators}
\newline
In the spin-2 case the parametrization of the two possible operators is
straightforward:
\begin{eqnarray}
\label{anh7}
\langle p S | \overline q_f (0)  
  \left[
      (  P_\mu\gamma_\nu  + P_\nu\gamma_\mu)  
        - \frac{1}{2} g_{\mu \nu} \Pslash \right]        
q_f (0) | p S \rangle &=& 2 a_f^{(1,2)} 
     \left [ 2p_\mu p_\nu - \frac{1}{2} g_{\mu \nu} M^2 \right]
\;, \\
\label{anh8}
\langle p S | \overline q_f (0)  
       (  P_\mu\gamma_\nu  + P_\nu\gamma_\mu)   \gamma_5
q_f (0) | p S \rangle &=& 2 a_{f5}^{(1,2)}
     \left [ p_\mu S_\nu + p_\nu S_\mu \right] \;.
\end{eqnarray}
For spin-1 only the pseudo-tensor operator yields a
non-zero matrix element:
\begin{eqnarray}
\label{anh9}
\langle p S | \overline q_f (0)
      (  P_\mu\gamma_\nu  - P_\nu\gamma_\mu)
q_f (0) | p S \rangle &=& 0
\;, \\
\label{anh10}
\langle p S | \overline q_f (0)
       (  P_\mu\gamma_\nu  - P_\nu\gamma_\mu)   \gamma_5
q_f (0) | p S \rangle &=&  m_f
\langle p S | \overline q_f (0)
  (-  i \sigma_{\mu \nu})   \gamma_5 
q_f (0) | p S \rangle 
\nonumber \\
&=& 2 \frac{m_f}{M} a_{f5-}^{(0,2)}  
\left [ p_\mu S_\nu - p_\nu S_\mu \right]
\;.
\end{eqnarray}
$a_{f5-}^{(0,2)}$ is 
connected to the first moment of 
the parton distribution $h_1(x)$ which is proportional to the structure 
function $G_3(x)$.
\newpage
\noindent
\underline{Rank-three tensor operators}
\newline
The operators $ P_\alpha P_\beta \gamma_\sigma$ and
$\gamma_\alpha \gamma_\beta P_\sigma$ have to be considered. 
\newline
{\bf $\gamma_\alpha \gamma_\beta P_\sigma$: Spin-2}
\newline
Again, only the operator with $\gamma_5$ has a nonvanishing matrix
element, i.e.:
\begin{eqnarray}
\label{anh12}
&&\langle p S | \left. \overline q_f (0) 
 \gamma_\alpha \gamma_\beta  P_\sigma \right|_{\rm spin-2} 
 q_f (0) |p S \rangle  =  0 \quad,
\\ \nonumber \\
&&\langle p S | \left. \overline q_f (0)
 \gamma_\alpha \gamma_\beta  P_\sigma \right|_{\rm spin-2} \gamma_5
 q_f (0) |p S \rangle  =  
\nonumber \\
&=& \langle p S | \overline q_f (0) \Bigg[ (-i) \sigma_{\alpha \beta} P^\sigma \gamma_5 
- \frac{1}{6} \left[
  g_{\alpha \sigma}  (\Pslash \gamma_\beta  - \gamma_\beta  \Pslash)
 -g_{\beta  \sigma}  (\Pslash \gamma_\alpha - \gamma_\alpha  \Pslash) \right] \gamma_5 \Bigg] 
 q_f (0) |p S \rangle
\nonumber \\
&=&  \frac{2}{M}a_{f5-}^{(1,2)} 
\left[ (p_\alpha S_\beta - p_\beta S_\alpha) p_\sigma 
      - \frac{M^2}{3} [ g_{\alpha \sigma} S_\beta 
                      - g_{\beta  \sigma} S_\alpha] \right] \quad.
\end{eqnarray}
$a_{f5-}^{(1,2)}$ is connected to $h_1(x)$ as indicated by  
$(-i) \sigma_{\alpha \beta}$.
\newline
\newline
{\bf $\gamma_\alpha \gamma_\beta P_\sigma$: Spin-0}
\newline
By means of the equation of motion we get:
\begin{eqnarray}
{1\over 6}\label{anh14}
\langle p S | \overline q_f (0) 
\epsilon_{\alpha \beta \sigma \delta} \epsilon^\delta_{\;\tau \rho \lambda} 
         \gamma^\tau \gamma^\rho P^\lambda 
q_f (0) |p S \rangle
& = & {1\over 3}im_f \epsilon_{\alpha \beta \sigma \delta} 
\langle p S | \overline q_f (0) \gamma^\delta \gamma_5 q_f (0)|p S\rangle
\nonumber \\
&=& {2\over 3}im_f \epsilon_{\alpha \beta \sigma \delta} a_{f5}^{(0,2)} S^\delta
\quad,
\\  
\label{anh14a}
{1\over 6}\langle p S | \overline q_f (0)
\epsilon_{\alpha \beta \sigma \delta} \epsilon^\delta_{\;\tau \rho \lambda}
         \gamma^\tau \gamma^\rho P^\lambda \gamma_5
q_f (0) |p S \rangle &=& 0 \quad.
\end{eqnarray}
\newline
{\bf $ P_\alpha P_\beta \gamma_\sigma$: Spin-3}
\newline
The parametrization of the spin-3 part is straightforward:
\begin{eqnarray}
\label{anh15}  
&&\langle p S | \left. \overline q_f (0)
P_\alpha P_\beta  \gamma_\sigma \right|_{\rm spin-3}
q_f (0) | p S \rangle 
 \nonumber \\
&=& 2 a_f^{(2,2)} \Bigg[ p_\alpha p_\beta p_\sigma 
- \frac{M^2}{6} [   g_{\alpha \beta  } p_\sigma
                  + g_{\beta  \sigma } p_\alpha
                  + g_{\alpha \sigma } p_\beta] \Bigg] \quad,
\\ \nonumber \\
\label{anh16}
&&\langle p S | \left. \overline q_f (0)
P_\alpha P_\beta  \gamma_\sigma \right|_{\rm spin-3}\gamma_5
q_f (0) | p S \rangle
 \nonumber \\
&=& 2 a_{f5}^{(2,2)} \left[ \frac{1}{3}
(    p_\alpha p_\beta   S_\sigma 
   + p_\beta  p_\sigma  S_\alpha  
   + p_\sigma p_\alpha  S_\beta   ) 
\right.
\nonumber \\
&& \left.\quad - \frac{M^2}{18} [  g_{\alpha \beta  } S_\sigma
                                 + g_{\beta  \sigma } S_\alpha
                                 + g_{\alpha \sigma } S_\beta] \right] \quad.
\end{eqnarray}
\newline
\newline
{\bf $ P_\alpha P_\beta \gamma_\sigma$: Spin-2}
\newline
In the spin-2 contribution the gluonic field strength tensor comes into play
due to the following two identities:
\begin{eqnarray}
\label{anh17}
g \tilde G^{\beta \sigma} &=&
\Bigg\{
  \Pslash ( P^\beta \gamma^\sigma - P^\sigma \gamma^\beta)
 -        ( P^\beta \gamma^\sigma - P^\sigma \gamma^\beta) \Pslash
\nonumber \\
&&+ \frac{1}{2} \left(
  \Pslash [\gamma^\beta \gamma^\sigma - \gamma^\sigma \gamma^\beta]\Pslash 
- P^2 [\gamma^\beta \gamma^\sigma - \gamma^\sigma \gamma^\beta] \right)
 \nonumber \\
&& + ( P^\beta P^\sigma - P^\sigma P^\beta) \Bigg\}\gamma_5 \quad,
\end{eqnarray}
and
\begin{eqnarray}
\label{anh18}
&&\gamma^\alpha g \tilde G^{\beta \sigma} +
\gamma^\beta  g \tilde G^{\alpha \sigma} 
+ \frac{1}{3}
 \left[ 2g^{\alpha \beta} g \tilde G^{\sigma \delta} \gamma_\delta
        -g^{\beta\sigma}g \tilde G^{\alpha\delta}\gamma_\delta
        -g^{\sigma\alpha}g \tilde G^{\beta\delta}\gamma_\delta
 \right]
\nonumber \\
&=&
\Bigg\{
\left( 2 P^\alpha P^\beta  \gamma^\sigma + 2 P^\beta  P^\alpha \gamma^\sigma
      -  P^\beta  P^\sigma \gamma^\alpha -   P^\sigma P^\beta  \gamma^\alpha
      -  P^\sigma P^\alpha \gamma^\beta  -   P^\alpha P^\sigma \gamma^\beta
\right.
\nonumber \\
&&- \frac{2}{3}\left.  P^2 \left[ 2g^{\alpha  \beta} \gamma^\sigma
                         -g^{\beta \sigma } \gamma^\alpha
                         -g^{\sigma \alpha} \gamma^\beta \right] 
\right) 
\nonumber \\
\nonumber \\
&& -m \bigg[
     \frac{1}{3} \left(
           2 \gamma^\beta   \gamma^\sigma  P^\alpha 
         -   \gamma^\sigma  \gamma^\alpha  P^\beta
         -   \gamma^\alpha  \gamma^\beta   P^\sigma 
         - 2 \gamma^\sigma  \gamma^\beta   P^\alpha
         +   \gamma^\alpha  \gamma^\sigma  P^\beta
         +   \gamma^\beta   \gamma^\alpha  P^\sigma 
                  \right)
\nonumber \\
&& \quad   - \frac{1}{3} \left( 
   g^{\alpha \beta }[\Pslash \gamma^\sigma - \gamma^\sigma \Pslash]               
-  g^{\alpha \sigma}[\Pslash \gamma^\beta  - \gamma^\beta  \Pslash]
    \right)
\nonumber \\
&&   \quad +  \frac{1}{3} \left(
           2 \gamma^\alpha   \gamma^\sigma   P^\beta
         -   \gamma^\sigma   \gamma^\beta    P^\alpha
         -   \gamma^\beta    \gamma^\alpha   P^\sigma
         - 2 \gamma^\sigma    \gamma^\alpha  P^\beta
         +   \gamma^\beta     \gamma^\sigma  P^\alpha
         +   \gamma^\alpha    \gamma^\beta   P^\sigma
                  \right)
\nonumber \\
&&  \quad  - \frac{1}{3} \left(
   g^{\beta \alpha }[\Pslash \gamma^\sigma - \gamma^\sigma \Pslash]       
-  g^{\beta \sigma}[\Pslash \gamma^\alpha  - \gamma^\alpha  \Pslash]
    \right) \bigg] \Bigg\} \gamma_5 \quad.
\end{eqnarray}
Again, one of the two possible matrix elements is zero
\begin{eqnarray}
\label{anh19}
&&\langle p S | \left. \overline q_f (0)
P_\alpha P_\beta  \gamma_\sigma \right|_{\rm spin-2}
q_f (0) | p S \rangle
\nonumber \\
& =&
\langle p S | \overline q_f (0) \frac{1}{6}
 \Bigg[ (\gamma_\alpha g \tilde G_{\beta \sigma} +
         \gamma_\beta  g \tilde G_{\alpha \sigma}) \gamma_5   
\nonumber \\
&&+ \frac{1}{3}
 \left[ 2g_{\alpha \beta} g \tilde G_{\sigma \delta} \gamma^\delta
        -g_{\beta\sigma}  g \tilde G_{\alpha \delta} \gamma^\delta
        -g_{\sigma\alpha} g \tilde G_{\beta  \delta} \gamma^\delta
 \right] \gamma_5 \Bigg]
q_f (0) | p S \rangle
 \nonumber \\
&=  & 0 \quad. 
\end{eqnarray}      
This implies that the gluonic matrix element is zero as well. In the other case
we have for the gluonic representation and for the representation in
terms of covariant derivatives:
\begin{eqnarray}
\label{anh20}
&&\langle p S | \overline q_f (0) \frac{1}{6}
 \Bigg[ (\gamma_\alpha g \tilde G_{\beta \sigma} +
\gamma_\beta  g \tilde G_{\alpha \sigma})
\nonumber \\
&&+ \frac{1}{3}
 \left[ 2g_{\alpha \beta} g \tilde G_{\sigma \delta} \gamma^\delta
        -g_{\beta\sigma}  g \tilde G_{\alpha \delta} \gamma^\delta
        -g_{\sigma\alpha} g \tilde G_{\beta  \delta} \gamma^\delta
 \right]  \Bigg]
q_f (0) | p S \rangle
\nonumber \\
&=& 2     a_{f5}^{(2,3)} \Bigg[\frac{1}{3} 
                            (2   p_\alpha p_\beta  S_\sigma
                               - p_\beta  p_\sigma S_\alpha
                               - p_\sigma p_\alpha S_\beta)
           - \frac{M^2}{9} [2 g_{\alpha \beta  } S_\sigma
                            - g_{\beta  \sigma } S_\alpha
                            - g_{\alpha \sigma } S_\beta] \Bigg] \quad,
\end{eqnarray}
and
\begin{eqnarray}
\label{anh21}
&&\langle p S | \left. \overline q_f (0)
P_\alpha P_\beta  \gamma_\sigma \right|_{\rm spin-2}\gamma_5
q_f (0) | p S \rangle
 \nonumber \\
&=& 2 \tilde a_{f5}^{(2,3)} \Bigg[\frac{1}{3} (2   p_\alpha p_\beta  S_\sigma 
                                                 - p_\beta  p_\sigma S_\alpha
                                                 - p_\sigma p_\alpha S_\beta)   
          - \frac{M^2}{9} [2 g_{\alpha \beta  } S_\sigma
                            - g_{\beta  \sigma } S_\alpha
                            - g_{\alpha \sigma } S_\beta]\Bigg]\;. 
\nonumber \\
\end{eqnarray}
These two representations are equivalent
except for mass terms. Explicitly one has the identity
\begin{equation}
\label{anh22a}
a_{f5}^{(2,3)} =  \tilde a_{f5}^{(2,3)}
-  \frac{m_f}{M} a_{f5-}^{(1,2)}
\quad.
\end{equation}
\newline
\newline
{\bf $ P_\alpha P_\beta \gamma_\sigma$: Spin-0}
\newline
For the spin-0 contribution there exists also a transformation 
between a gluonic representation
and a representation in form of covariant derivatives only:
\begin{eqnarray}
\label{anh23}   
\langle p S | \overline q_f (0) 
   \epsilon_{\alpha \beta \sigma \delta} 
\epsilon^\delta_{ \; \tau \rho \lambda}
 P^\tau P^\rho \gamma^\lambda q_f (0) |  p S \rangle 
&=&
i \langle p S |  \overline q_f (0) \epsilon_{\alpha \beta \sigma \delta}
                      g \tilde    G^\delta_{\; \lambda} \gamma^\lambda 
                                  q_f (0) |  p S \rangle
\nonumber \\
&=& 2i  \epsilon_{\alpha \beta \sigma \delta} M^2  a_{f5}^{(2,4)} S^\delta 
\nonumber \\
&=& i\epsilon_{\alpha \beta \sigma \delta} 
\langle p S |  \overline q_f (0) (P^2 - m_f^2) \gamma^\delta \gamma_5 q_f (0) |  p S \rangle
\nonumber \\
&=& 2i\epsilon_{\alpha \beta \sigma \delta}
     \left[ M^2 \tilde a_{f5}^{(2,4)} -m_f^2 a_{f5}^{(0,2)} \right] S^\delta \;,
\end{eqnarray}
and analogously
\begin{eqnarray}
\label{anh24}   
\langle p S | \overline q_f (0)
   \epsilon_{\alpha \beta \sigma \delta} 
\epsilon^\delta_{ \; \tau \rho \lambda}
 P^\tau P^\rho \gamma^\lambda \gamma_5 q_f (0) |  p S \rangle
&=&
i \langle p S |  \overline q_f (0) \epsilon_{\alpha \beta \sigma \delta}
                     g  \tilde   G^\delta_{\; \lambda} \gamma^\lambda 
     \gamma_5 q_f (0) |  p S \rangle
\nonumber \\
&=& 2i  \epsilon_{\alpha \beta \sigma \delta} M^2  a_{f}^{(2,4)} P^\delta
\nonumber \\
&=& i\epsilon_{\alpha \beta \sigma \delta}
\langle p S |  \overline q_f (0) (P^2 - m_f^2) \gamma^\delta q_f (0) |  p S \rangle
\nonumber \\
&=& 2i\epsilon_{\alpha \beta \sigma \delta}
     \left[ M^2 \tilde a_{f}^{(2,4)} -m_f^2 a_{f}^{(0,2)} \right] P^\delta \;.
\end{eqnarray}
Obviously
\begin{eqnarray}
\label{anh25}   
M^2 a_{f5}^{(2,4)} = M^2 \tilde a_{f5}^{(2,4)}  -m_f^2 a_{f5}^{(0,2)} \quad,\\
M^2 a_{f}^{(2,4)}  = M^2 \tilde a_{f} ^{(2,4)}  -m_f^2 a_{f}^{(0,2)}  \quad.
\end{eqnarray} 
\section{Operators generated by the cat ear diagram}
\label{appB}
\noindent

In the required approximation the cat ear diagram (fig.4)  
is not connected with    
the external gluonic field, so it is sufficient to analyze ordinary
Feynman diagrams in zeroth order in $(\Pslash - m_f)/Q$.
In the case of $\gamma$ or $Z^0$ exchange the corresponding 
virtual forward Compton scattering amplitude then is given by
\begin{eqnarray}
T_{\mu \nu}^{({\rm cat\; ear}\; \gamma, Z^0)} \Bigg|_{0.\; order}
& =&
\sum_{ff'} \frac{g^2}{2!} i^3 \int d^4 \xi e^{iq\xi} \int d z_1 dz_2
\langle pS| \bigg\{ 
        \overline q_f (\xi)\gamma_\mu (V_f + A_f \gamma_5) 
                       iS_f(\xi,z_1)\gamma^\rho t^a q_f(z_1)
\nonumber \\ && \qquad
      +  \overline q_f (z_1)\gamma^\rho t^a iS_f(z_1,\xi)
                       \gamma_\mu  (V_f + A_f \gamma_5) q_f(\xi)\bigg\}
     i D_{\rho \sigma}^{ab} (z_1, z_2)
\nonumber \\ && \qquad \times\bigg\{ 
        \overline q_{f'} (z_2)\gamma^\sigma t^b iS_{f'}(z_2,0)
                 \gamma_\nu  (V_{f'} + A_{f'} \gamma_5) q_{f'}(0)
\nonumber \\ && \qquad
   + \overline q_{f'} (0)\gamma_\nu (V_{f'} + A_{f'} \gamma_5)
                       iS_{f'}(0,z_2)\gamma^\sigma t^b q_{f'}(z_2)\bigg\}
|pS \rangle \quad,
\end{eqnarray}
where we have to insert the free massless fermion-propagator.
$D^{ab}_{\rho \sigma}(z_1,z_2)$ is the
gluonic propagator connecting the ``ears'' of the cat ear diagram.
$t_a$, $a = 1,\dots, 8$ denote the color matrices 
($\rm{tr} (t^a t^b) =  \delta^{ab}/2$). 
In that way we obtain
\begin{eqnarray}
T_{\mu \nu}^{(\rm cat\; ear \; Z^0, \gamma)}\Bigg|_{0.\; order} &=&
-\frac{g^2}{2} \sum_{ff'}
\langle pS| \bigg\{\overline q_f (0) \gamma_\mu (V_f+ A_f\gamma_5) 
            \frac{1}{\qslash}
            \gamma_\rho t^a q_f(0) 
\nonumber \\ && \quad -
             \overline q_f (0) \gamma_\rho t^a 
             \frac{1}{\qslash}
             \gamma_\mu (V_f+ A_f\gamma_5) q_f(0) \bigg\} 
\frac{-g^{\sigma \rho}}{q^2}
\nonumber \\  && \quad \times \bigg\{
             \overline q_{f'} (0)\gamma_\sigma t^a  
             \frac{1}{\qslash}
             \gamma_\nu (V_{f'}+ A_{f'}\gamma_5)q_{f'}(0)              
\nonumber \\ && \quad -
            \overline q_{f'} (0)\gamma_\nu (V_{f'}+ A_{f'}\gamma_5)
            \frac{1}{\qslash}
            \gamma_\sigma t^a q_{f'} (0) \bigg\}|pS \rangle \quad.
\end{eqnarray}
The corresponding expression for the $W^-$ exchange is
\begin{eqnarray}
T_{\mu \nu}^{(\rm cat\; ear \; W^-)} \Bigg|_{0.\; order} & =&
-\frac{g^2}{2} 
\langle pS| \bigg\{\overline u (0) \gamma_\mu (V+ A\gamma_5) 
            \frac{1}{\qslash}
            \gamma_\rho t^a d(0) 
\nonumber \\ && \quad -
             \overline u (0) \gamma_\rho t^a 
             \frac{1}{\qslash}
             \gamma_\mu (V+ A\gamma_5) d (0) \bigg\} 
\frac{-g^{\sigma \rho}}{q^2}
\nonumber \\  && \quad \times \bigg\{
             \overline d (0)\gamma_\sigma t^a  
             \frac{1}{\qslash}
             \gamma_\nu (V + A\gamma_5)u(0)              
\nonumber \\ && \quad -
            \overline d (0)\gamma_\nu (V+ A\gamma_5)
            \frac{1}{\qslash}
            \gamma_\sigma t^a u (0) \bigg\}|pS \rangle
\nonumber \\
&& + \;{\rm further\; flavor\; combinations } \quad.
\end{eqnarray}
(The case of $W^+$ exchange is obtained by exchanging the quark flavors.) 
Introducing reduced matrix elements of the occurring four-quark operators
we finally get in zeroth order 
\begin{eqnarray}
\label{53}
T_{\mu \nu}^{(\rm cat\; ear \; Z^0 , \gamma)}\bigg|_{0.\; {\rm order}}
&=&
g^2 \Bigg\{ 
    \left(- g_{\mu \nu} + \frac{q_\mu q_\nu}{q^2} \right)
    \left[ \frac{\omega^2}{Q^2} C_=^{(2)} 
           + 2 \frac{M^2}{Q^4} \left[ C_=^{(0)} + C_=^{(2)} \right]
    \right] 
\nonumber \\
\nonumber \\
&&  + \frac{Sq}{\nu} 
    \left(- g_{\mu \nu} + \frac{q_\mu q_\nu}{q^2} \right)
    \left[ \frac{\omega^2}{Q^2} C_\times^{(2)} \right]
\nonumber \\
\nonumber \\
&& + \frac{1}{\nu} \hat p_\mu \hat p_\nu 
    \left[ 2 \frac{\omega}{Q^2} C_=^{(2)} \right]
 + \frac{1}{2\nu} ( \hat S_\mu \hat p_\nu + \hat S_\nu \hat p_\mu)
     \left[ 2 \frac{\omega}{Q^2} C_\times^{(2)} \right] \quad,
\nonumber \\
\nonumber \\
\nonumber \\
T_{\mu \nu}^{(\rm cat\;ear \;W^- )}\Bigg|_{0.\; {\rm order}}
&=&
g^2 \Bigg\{ 
    \left(- g_{\mu \nu} + \frac{q_\mu q_\nu}{q^2} \right)
    \left[ \frac{\omega^2}{Q^2} C_{ud=}^{(2)} 
           + 2 \frac{M^2}{Q^4} \left[ C_{ud=}^{(0)} + C_{ud=}^{(2)} \right]
    \right] 
\nonumber \\
\nonumber \\
&&  + \frac{Sq}{\nu} 
    \left(- g_{\mu \nu} + \frac{q_\mu q_\nu}{q^2} \right)
    \left[ \frac{\omega^2}{Q^2} C_{ud \times}^{(2)} \right]
\nonumber \\
\nonumber \\
&& + \frac{1}{\nu} \hat p_\mu \hat p_\nu 
    \left[ 2 \frac{\omega}{Q^2} C_{ud=}^{(2)} \right]
 + \frac{1}{2\nu} ( \hat S_\mu \hat p_\nu + \hat S_\nu \hat p_\mu)
     \left[ 2 \frac{\omega}{Q^2} C_{ud \times}^{(2)} \right] \quad,
\end{eqnarray} 
where only the expression for the $u \rightarrow d$ flavor
combination is given.
Here the flavor independent matrix elements $C_{= / \times}^{(j)}$ are defined
by 
\begin{equation}
C_{= / \times}^{(j)} := \sum_f C_{ff = / \times}^{(j)}, \qquad j=0,1,2,\dots \quad.
\nonumber \end{equation}
Note, that the cat ear diagram (\ref{53}) 
is symmetric in $\mu$ and $\nu$. 
Therefore in zeroth order of $\Pslash - m_f$ 
it gives only twist-4 and 
twist-6 contributions to structure functions 
associated with symmetric Lorentz structures, such as $F_1$ and $F_2$. 
Since in the required approximation quark masses do not occur, the
cat ear does not contribute to structure functions given by non-conserved 
currents. Therefore we find that cat-ear diagrams
contribute only 
to the structure functions 
$F_1,F_2,a_1$, and $b_2$. 

For the definition of the reduced matrix elements of the cat ear diagram
we need the following definition of four-quark operators:
\begin{eqnarray}
\hat {\cal O}_{55,ff'}^{\sigma \sigma'} &:= &
\overline  q_{f }  (0) \gamma^\sigma  \gamma_5 t^a q_{f'}(0)
\overline  q_{f'}(0) \gamma^{\sigma'} \gamma_5 t^a q_{f}(0)A_f A_{f'} \quad,
\\
\hat {\cal O}_{5X,ff'}^{\sigma \sigma'} &:= &
\overline  q_{f }  (0) \gamma^\sigma  \gamma_5 t^a  q_{f'}(0) 
\overline  q_{f'}(0) \gamma^{\sigma'} t^a   q_{f}(0) A_f V_{f'} \quad,
\\
\hat {\cal O}_{X5,ff'}^{\sigma \sigma'} &:= &
\overline  q_{f }  (0) \gamma^\sigma  t^a   q_{f'}(0)
\overline  q_{f'}(0) \gamma^{\sigma'} \gamma_5 t^a  q_{f}(0) V_f A_{f'} \quad,
\\
\hat {\cal O}_{XX,ff'}^{\sigma \sigma'} &:= &
\overline  q_{f }  (0) \gamma^\sigma t^a    q_{f'}(0)
\overline  q_{f'}  (0) \gamma^{\sigma'}  t^a  q_{f }(0) V_{f} V_{f'} \quad.
\end{eqnarray}
Note that for $\gamma$ and $Z^0$ exchange $f = f'$, 
and in the case of $W^\pm$ exchange we always have
$V_f = V_{f'} = V$ and $A_f = A_{f'} = A$. 
We then define the reduced matrix elements to be
\begin{eqnarray}
2C^{(2)}_{ff'=} 
\left( p^\sigma p^{\sigma'} - \frac{M^2}{4}g^{\sigma \sigma'} \right)
&=& \langle pS | 
     \frac{1}{2} (\hat {\cal O}_{55,ff'}^{\sigma \sigma'}
                 +\hat {\cal O}_{XX,ff'}^{\sigma \sigma'}
                 +\hat {\cal O}_{55,ff'}^{\sigma' \sigma}
                 +\hat {\cal O}_{XX,ff'}^{\sigma' \sigma})
\nonumber \\
&& \quad 
    -\frac{1}{4} g^{\sigma \sigma'}  g_{\alpha \alpha'}
                ( \hat {\cal O}_{55,ff'}^{\alpha \alpha'}
                 +\hat {\cal O}_{XX,ff'}^{\alpha \alpha'}) 
    |pS \rangle \quad,
\\
\nonumber \\
0 &=& \langle pS |
     \frac{1}{2} (\hat {\cal O}_{55,ff'}^{\sigma \sigma'}
                 +\hat {\cal O}_{XX,ff'}^{\sigma \sigma'}
                 -\hat {\cal O}_{55,ff'}^{\sigma' \sigma} 
                 -\hat {\cal O}_{XX,ff'}^{\sigma' \sigma})
|pS \rangle \quad,
\\
\nonumber \\
2 M^2 C^{(0)}_{ff'=}
&=& \langle pS |
                           g_{\alpha \alpha'}  
                ( \hat {\cal O}_{55,ff'}^{\alpha \alpha'}
                 +\hat {\cal O}_{XX,ff'}^{\alpha \alpha'})
    |pS \rangle \quad,
\\
\nonumber \\
\nonumber \\
C^{(2)}_{ff' \times}
\left( S^\sigma p^{\sigma'} + S^{\sigma'}p^\sigma \right)
&=& \langle pS |
     \frac{1}{2} (\hat {\cal O}_{5X,ff'}^{\sigma \sigma'} 
                 +\hat {\cal O}_{X5,ff'}^{\sigma \sigma'} 
                 +\hat {\cal O}_{5X,ff'}^{\sigma' \sigma} 
                 +\hat {\cal O}_{X5,ff'}^{\sigma' \sigma})
\nonumber \\
&& \quad
    -\frac{1}{4} g^{\sigma \sigma'}  g_{\alpha \alpha'}  
                ( \hat {\cal O}_{5X,ff'}^{\alpha \alpha'}
                 +\hat {\cal O}_{X5,ff'}^{\alpha \alpha'})
    |pS \rangle \quad,
\\
\nonumber \\
0&=& \langle pS |
     \frac{1}{2} (\hat {\cal O}_{5X,ff'}^{\sigma \sigma'} 
                 +\hat {\cal O}_{X5,ff'}^{\sigma \sigma'} 
                 -\hat {\cal O}_{5X,ff'}^{\sigma' \sigma} 
                 -\hat {\cal O}_{X5,ff'}^{\sigma' \sigma}) \quad,
\\
\nonumber \\
0&=& \langle pS |
                 g_{\alpha \alpha'}   
                ( \hat {\cal O}_{5X,ff'}^{\alpha \alpha'}
                 +\hat {\cal O}_{X5,ff'}^{\alpha \alpha'})
    |pS \rangle \quad.
\end{eqnarray}
Note, that the operators $ \hat {\cal O}_{XX,ff'}, \hat {\cal O}_{5X,ff'},
\hat {\cal O}_{X5,ff'}$
can be transformed by means of the equation of motion  
\begin{equation}
D^\alpha G_{\alpha \beta}^a 
= - \sum_f  g \overline q_f \gamma_\beta t^a q_f,
\end{equation}
where $D_\alpha$ is the covariant QCD derivative and $G_{\alpha \beta}^a$
the gluonic field tensor. However, as this operation cannot be applied to 
$ \hat {\cal O}_{55,ff'}$, we did not perform it to keep a uniform notation. 
\end{appendix}

\newpage
{\LARGE  Figure Captions}
\newline
\newline
{\bf Figure~1:} Kinematic variables in deep inelastic scattering:
An electron with four-momentum $k$ and spin $\sigma$ 
is scattered by a nucleon 
with four-momentum $p$ and polarization $\lambda$ 
through the exchange of a gauge boson
with four-momentum $q$. The nucleon fragments into a variety of hadrons $X$,
which together carry the four-momentum $p_X$.
\newline \newline \newline
{\bf Figure~2:} The contour of integration of the virtual forward Compton
scattering amplitude $T_{\mu \nu}$. 
The regions where the function $T_{\mu \nu}(\omega)$ has singularities 
are shadowed. 
The four big arrows show the integration
path. It can be completed by the outer circle in the infinity thus
being equivalent to the inner circle. 
\newline \newline  \newline
{\bf Figure~3:} The hand bag diagram: It represents the virtual 
Compton forward scattering amplitude, for quark (solid line)
gauge boson (sinuous line) scattering. 
\newline \newline  \newline
{\bf Figure~4:} The cat ear diagram:
The lines represent: Gauge boson (sinuous line), gluon (curly line), 
quark (straight line).
\newpage
\begin{table}
\begin{center}
\begin{tabular}{c | c |c}
 &  unpolarized structure functions &  polarized structure functions 
\\
\hline
\underline{parity-conserved:}  & &
\\
current conserved &  $F_1$, $F_2$ & $g_1$, $g_2$  
\\
current non-conserved & $F_4$, $F_5$ & $ g_3 $
\\
\hline
\underline{parity-violating:}  & &
\\
current conserved  &  $F_3$  & $a_1$, $a_2$, $b_1$
\\
current non-conserved &  &$a_4$, $a_5$, $b_2$ 
\\
\end{tabular}
\end{center} 
\caption{The 14 nucleon structure functions.}
\label{tab1}
\end{table}
\begin{table} 
\begin{center}
\begin{tabular}{r c l }
\multicolumn{3}{c}{Moments of structure functions} \\
\hline
\hline
$\int_0^1 x dx F^{(\gamma, Z^0)}_1(x,Q^2) $&=&$ \sum_{f} \frac{1}{2} a_f^{(1,2)}
   \left( V_f^2 +  A_f^2 \right) + {\cal O}(1/Q^2)$ \\
$\int_0^1 dx F^{(\gamma, Z^0)}_2(x,Q^2) $&=&$  \sum_{f} a_f^{(1,2)}
   \left( V_{f}^2 +  A_{f}^2 \right) + {\cal O}(1/Q^2) $\\   
&&\\
$\int_0^1 dx F^{(\gamma, Z^0)}_3(x,Q^2) $&=&$
\sum_{f}\left\{ a_f^{(0,2)} +
\frac{2}{9} \frac{M^2}{Q^2} \left[ 3 a_{f}^{(2,2)}
                  +  8  a_{f}^{(2,4)} \right] \right.$\\
&& $\qquad   \left.
- \frac{2}{3} \frac{m_f^2}{Q^2} a_{f}^{(0,2)}
  \right\} 2   V_f  A_{f} + {\cal O} (1/Q^4)$ \\
$\int_0^1 x^2 dx F^{(\gamma, Z^0)}_3 (x,Q^2)  $&=&$ \sum_f
 a_{f}^{(2,2)} 2V_f A_f    + {\cal O} (1/Q^2) $ \\  
&&\\
$\int_0^1 dx F^{(\gamma, Z^0)}_4(x,Q^2) $&=&$ 
\sum_f \frac{m_f M}{Q^2} a_{f-}^{(0,3)} A_f^2 + {\cal O} (1/Q^4)$\\          
$\int_0^1 x^2dx F^{(\gamma, Z^0)}_4(x,Q^2) $&=&$ 0 + {\cal O} (1/Q^2)$\\
&&\\
$\int_0^1 xdx F^{(\gamma, Z^0)}_5(x,Q^2) $&=&$ 0 + {\cal O} (1/Q^2)$\\
&&\\
&&\\
$\int_0^1 dxF_1^{(\overline \nu - \nu)}(x,Q^2) $&=&$
  a_{\Delta V}^{(0,2)}
       + \frac{8}{9} \frac{M^2}{Q^2} [3a_{\Delta V}^{(2,2)}
                                    -  a_{\Delta V}^{(2,4)}]$ \\ 
&&$       + \frac{1}{3} \frac{1}{Q^2}[m_{\Delta V}^2
                            - 3 {m'}_{\Delta V}^2]a_{\Delta V}^{(0,2)}
     +{\cal O}(1/Q^4)$\\

$\int_0^1 x dx F^{(\overline \nu + \nu)}_1(x,Q^2) $&=&$
 a_S^{(1,2)} + {\cal O}(1/Q^2)$\\
$\int_0^1 x^2dx F^{(\overline \nu - \nu)}_1(x,Q^2) $&=&$
a_{\Delta V}^{(2,2)} +{\cal O}(1/Q^2)$\\
&&\\
$\int_0^1 dx F_2^{(\overline \nu + \nu)}(x,Q^2) $&=&$
2a_S^{(1,2)} + {\cal O}(1/Q^2)$\\
$\int_0^1 x dx  F^{(\overline \nu - \nu)}_2(x,Q^2) $&=&$
2 a_{\Delta V}^{(2,2)} +{\cal O}(1/Q^2)$\\
&&\\
$\int_0^1 dx F_3^{(\overline \nu + \nu)}(x,Q^2)
$&=&$ -2 a_V^{(0,2)}
-\frac{4}{9} \frac{M^2}{Q^2} [3a_V^{(2,2)} + 8 a_V^{(2,4)}]$\\
&& $ -\frac{2}{3} \frac{1}{Q^2}
                [m_V^2 - 3 {m'}_V^2] a_V^{(0,2)}
+{\cal O}(1/Q^4)$\\ 
$\int_0^1x dx F_3^{(\overline \nu - \nu)}(x,Q^2)
$&=&$ -2a_{\Delta S}^{(1,2)} + {\cal O}(1/Q^2)$\\
$\int_0^1 x^2 dx F_3^{(\overline \nu + \nu)}(x,Q^2) $&=&$
-2 a_V^{(2,2)}+ {\cal O}(1/Q^2)$\\ 
&&\\
$\int_0^1 dx  F_4^{(\overline \nu + \nu)}(x,Q^2) $&=&$
 \frac{m_S M}{Q^2}
 a_{S-}^{(0,3)}+ {\cal O}(1/Q^4)$\\
$\int_0^1 xdx F_4^{(\overline \nu - \nu)}(x,Q^2) $&=&$
+ \frac{1}{2} \frac{1}{Q^2} [3m_{\Delta V}^2 - {m'}_{\Delta V}^2]
a_{\Delta V}^{(0,2)}
+ {\cal O}(1/Q^4)$\\
$\int x^2dx F^{(\overline \nu + \nu)}_4(x,Q^2) $&=&$ 0  + {\cal O}(1/Q^2)$\\
$\int x^3dx F^{(\overline \nu - \nu)}_4(x,Q^2) $&=&$ 0  + {\cal O}(1/Q^2)$\\
&&\\
$\int_0^1 dx F^{(\overline \nu - \nu)}_5(x,Q^2) $&=&$
+ 2\frac{1}{Q^2} [m_{\Delta V}^2 - {m'}_{\Delta V}^2]
a_{\Delta V}^{(0,2)} +{\cal O}(1/Q^4)$\\
$\int_0^1 xdx F^{(\overline \nu + \nu)}_5(x,Q^2) $&=&$ 0 + {\cal O}(1/Q^2)$\\
$\int_0^1 x^2dx F^{(\overline \nu - \nu)}_5(x,Q^2) $&=&$ 0 + {\cal O}(1/Q^2)$\\
&&\\
&&\\
$\int_0^1 dx g_1^{(\gamma, Z^0)}(x,Q^2) $&=&$ \sum_f \left[\left\{
\frac{1}{2} a_{f5}^{(0,2)}
+ \frac{1}{9} \frac{M^2}{Q^2}\left( a_{f5}^{(2,2)}
+ 4   a_{f5}^{(2,3)} + 4  a_{f5}^{(2,4)}
+ 4   \frac{m_f}{M}  a_{f5-}^{(1,2)}\right)
\right.  \right.$\\
&& $  \qquad \left.   - \frac{2}{9} \frac{m_f^2}{Q^2} a_{f5}^{(0,2)}
\right\}(V_f^2 + A_f^2)
$\\ && $\qquad \left.
-\left\{  \frac{1}{3} \frac{m_f^2}{Q^2} a_{f5}^{(0,2)}
             \right\} (V_f^2 - A_f^2) \right]
+ {\cal O} (1/Q^4)$\\
$\int_0^1 x^2dx g_1^{(\gamma, Z^0)}(x,Q^2) $&=&$ \sum_f
\frac{1}{2} a_{f5}^{(2,2)}
(V_f^2 + A_f^2) + {\cal O} (1/Q^2)$\\
&&\\
$\int_0^1 dx g_2^{(\gamma, Z^0)}(x,Q^2) $&=&$ 0 + {\cal O} (1/Q^4)$\\
$\int_0^1 x^2dx g_2^{(\gamma, Z^0)}(x,Q^2) $&=&$ -\sum_f
 \frac{1}{3} \left\{ a_{f5}^{(2,2)} -   a_{f5}^{(2,3)}
 - \frac{m_f}{M} a_{f5-}^{(1,2)}
\right\}
(V_f^2 + A_f^2) + {\cal O} (1/Q^2)$\\
&&\\
$\int_0^1 x dx g_3^{(\gamma, Z^0)}(x,Q^2)$& =&$ 0 + {\cal O} (1/Q^2)$\\
&&\\
$\int_0^1 dx g_1^{(\overline \nu + \nu)}(x,Q^2) $&=&$
 a_{S5}^{(0,2)}
+ \frac{2}{9} \frac{M^2}{Q^2}\left( a_{S5}^{(2,2)}
+ 4  a_{S5}^{(2,3)} + 4  a_{S5}^{(2,4)}
+  4 \frac{m_S}{M} a_{S5-}^{(1,2)}
\right) $ \\
&&\\
&&$+ \frac{1}{9} \frac{1}{Q^2}[5m_S^2 - 9{m'}_S^2] a_{S5}^{(0,2)}
+  {\cal O} (1/Q^4)$\\
$\int_0^1 xdx g^{(\overline \nu - \nu)}_1(x,Q^2) $&=&$
a_{\Delta V 5}^{(1,2)}  + {\cal O} (1/Q^2)$\\
$\int_0^1 x^2dx g^{(\overline \nu + \nu)}_1(x,Q^2) $&=&$
a_{S5}^{(2,2)} + {\cal O} (1/Q^2)$\\
&&\\
$\int_0^1 dx g^{(\overline \nu + \nu)}_2(x,Q^2) $&=&$ 0 + {\cal O} (1/Q^4)$\\
$\int_0^1 xdx g_2^{(\overline \nu - \nu)}(x,Q^2) $&=&$ - \frac{1}{2}
\Bigg(
a_{\Delta V5}^{(1,2)} - \frac{m_{\Delta V}}{M}a_{\Delta V 5-}^{(0,2)}
\Bigg)
 + {\cal O} (1/Q^2)$\\
$\int_0^1 x^2dx g^{(\overline \nu + \nu)}_2(x,Q^2) $&=&$ - \frac{2}{3}
\left[a_{S5}^{(2,2)}-  a_{S5}^{(2,3)}  
- \frac{m_S}{M} a_{S5-}^{(1,2)} \right]
+ {\cal O} (1/Q^2)$\\
&&\\
$\int_0^1 dx g_3^{(\overline \nu - \nu)} (x,Q^2) $&=&$
\frac{m_{\Delta V}}{M}a_{\Delta V5-}^{(0,2)}
+ {\cal O} (1/Q^2)$\\
$\int_0^1 x dx g_3^{(\overline \nu + \nu)} (x,Q^2) $&=&$ 0 
+ {\cal O} (1/Q^2)$\\
&&\\
&&\\
$\int_0^1 x dx a_1^{(\gamma, Z^0)}(x,Q^2) $&=&$ \sum_{f}  a_{f5}^{(1,2)}
                                A_f V_f+ {\cal O} (1/Q^2)$\\
$\int_0^1  dx a_2^{(\gamma, Z^0)}(x,Q^2) $&=&$ 0 + {\cal O} (1/Q^2)$\\
$\int_0^1 x^2 dx a_4^{(\gamma, Z^0)}(x,Q^2) $&=&$
- \sum_{f}\frac{1}{2}  \frac{m_f}{M}a_{f5-}^{(0,2)}
                        A_f V_f+ {\cal O} (1/Q^2)$\\
$\int_0^1 x dx a_5^{(\gamma, Z^0)}(x,Q^2) $&=&$
- \sum_f \frac{m_f}{M}a_{f5-}^{(0,2)}
                   A_f V_f + {\cal O} (1/Q^2)$\\
&&\\
$\int_0^1  dx b_1^{(\gamma, Z^0)}(x,Q^2) $& = &$2\sum_{f}
                    (a_{f5}^{(1,2)} + \frac{m_f}{M}a_{f5-}^{(0,2)})
                                A_f V_f + {\cal O} (1/Q^2) $\\
$\int_0^1  dx b_2^{(\gamma, Z^0)}(x,Q^2) $&=&$ -2\sum_{f}
\frac{m_f}{M} a_{f5-}^{(0,2)}
                            A_f V_f + {\cal O} (1/Q^2)$\\
&&\\
&&\\
$\int_0^1 dx a_1^{(\overline \nu - \nu)}
 (x,Q^2) $&=&$ - a_{\Delta S5}^{(0,2)}   
-\frac{8}{9} \frac{M^2}{Q^2} [a_{\Delta S5}^{(2,2)}
  - 2  a_{\Delta S5}^{(2,3)} +  a_{\Delta S5}^{(2,4)} ]
+ \frac{4}{9} \frac{m_{\Delta S} M}{Q^2} a_{\Delta S5-}^{(1,2)}
$\\ && $
      - \frac{1}{9} \frac{1}{Q^2}
[5m_{\Delta S}^2 - 9{m'}_{\Delta S}^2] a_{\Delta S5}^{(0,2)}
+ {\cal O} (1/Q^4)$\\
$\int_0^1 xdx a_1 ^{(\overline \nu + \nu)}(x,Q^2) $&=&$ -
a_{V5}^{(1,2)} + {\cal O} (1/Q^2)$\\
$\int_0^1 x^2dx a_1^{(\overline \nu - \nu)} (x,Q^2) $&=&$
- a_{\Delta S5}^{(2,2)}+ {\cal O} (1/Q^2)$\\
&&\\
$\int_0^1 dx a_2 ^{(\overline \nu + \nu)}(x,Q^2) $&=&$ 0 + {\cal O} (1/Q^2)$\\
$\int_0^1 x dx a_2^{(\overline \nu - \nu)} (x,Q^2) $&=&$
-\frac{2}{3}a_{\Delta S5}^{(2,2)} + \frac{8}{3}  a_{\Delta S5}^{(2,3)}
        + \frac{2}{3} \frac{m_{\Delta S}}{M} a_{\Delta S5-}^{(1,2)}   
+ {\cal O} (1/Q^2) $\\
&&\\
$\int_0^1 xdx a_4^{(\overline \nu - \nu)}(x,Q^2) $&=&$
        + \frac{1}{2} \frac{1}{Q^2}
[m_{\Delta S}^2 -{m'}_{\Delta S}^2]a_{\Delta S5}^{(0,2)}
+ {\cal O} (1/Q^4)$\\
$\int_0^1 x^2dx a_4 ^{(\overline \nu + \nu)}(x,Q^2) $&=&$
\frac{1}{2} \frac{m_V}{M}a_{V5-}^{(0,2)}
+ {\cal O} (1/Q^2)$\\
$\int_0^1 x^3dx a_4 ^{(\overline \nu - \nu)}(x,Q^2)$&=&$
 \frac{1}{2} \frac{ m_{\Delta S}}{M}
a_{\Delta S5-}^{(1,2)} + {\cal O} (1/Q^2)$\\
&&\\
$\int_0^1 dx a_5^{(\overline \nu - \nu)}(x,Q^2) $&=&$
 0 + {\cal O} (1/Q^4)$\\
$\int_0^1 xdx a_5^{(\overline \nu + \nu)}(x,Q^2)$&=&$
 \frac{m_V}{M}a_{V5-}^{(0,2)} + {\cal O} (1/Q^2)$\\
$\int_0^1 x^2dx a_5^{(\overline \nu - \nu)}(x,Q^2) $&=&$
\frac{m_{\Delta S}}{M}
 a_{\Delta S5-}^{(1,2)} + {\cal O} (1/Q^2)$\\
&&\\
$\int_0^1  dx b^{(\overline \nu + \nu)}_1(x,Q^2)
$&=&$ -2\bigg( a_{V5}^{(1,2)} + \frac{m_V}{M} a_{V5-}^{(0,2)} \bigg)
+ {\cal O} (1/Q^2)$\\
$\int_0^1 x dx b_1^{(\overline \nu - \nu)}(x,Q^2) $&=&$
  - \Bigg[      \frac{4}{3}  a_{\Delta S5}^{(2,2)}
               +\frac{8}{3}  a_{\Delta S5}^{(2,3)}
               +\frac{8}{3} \frac{m_{\Delta S}}{M}
               a_{\Delta S5-}^{(1,2)} \Bigg]
+ {\cal O} (1/Q^2)$\\
&&\\
$\int_0^1  dx b_2^{(\overline \nu + \nu)}(x,Q^2) $&=&$ 2
\frac{m_V}{M}a_{V5-}^{(0,2)}
+ {\cal O} (1/Q^2) $\\
$\int_0^1  xdx b_2^{(\overline \nu - \nu)}(x,Q^2) $&=&$
2\frac{ m_{\Delta S}}{M}  a_{\Delta S 5-}^{(1,2)}
+ {\cal O} (1/Q^2)$\\
\end{tabular}
\end{center}
\caption{Moments of structure functions with calculated $Q^2$ corrections.
Both the neutrino and the $\gamma/Z^0$ case are given.}
\label{tab4}
\end{table} 
\newpage
\begin{table}
\begin{center}
\begin{tabular}{l| c |  c }
& sum rule &  reference
\\
\hline
\hline
unpolarized && \\
&$\int_0^1 dx(F_1^{\overline \nu p} - F_1^{\nu p}) = 1 \eta_W$
& Bjorken (1967)  \cite{Bj67} \\  
\hline
&$\int_0^1 \frac{dx}{x} (F_2^{\overline \nu p} - F_2^{\nu p}) = 2\eta_W$
& Adler (1966)  \cite{Ad66}
\\
\hline 
&$\int_0^1 d x (F_3^{\overline \nu p} + F_3^{ \nu p}) = -6\eta_W$&
Gross, Llewellyn-Smith (1969) \cite{GLS69}
\\
\hline
\hline
polarized && \\
& $\int_0^1 dx (g_1^p(x) - g_1^n(x)) = \frac{C}{2} \frac{g_A}{g_V}$&
 Bjorken (1966)  \cite{Bj66}
\\
\hline
& $\int_0^1 dx g_1^p(x) =  \frac{1}{18} \frac{g_A}{g_V}
\frac{9F/D -1}{F/D+1}$&
 Ellis, Jaffe (1974)  \cite{EJ74}
\\
\hline
&$ \int_0^1 dx  a_1 ^{(\overline \nu - \nu)(p-n)}=  - 2   \frac{g_A}{g_V}
\eta_W$ &
 Wray (1972) \cite{Wr72} 
\\
\end{tabular}
\end{center}
\caption{Sum rules in the Bjorken limit ($Q^2 \to \infty$).}
\label{tab2}
\end{table}
\newpage
\begin{table}
\begin{center}
\begin{tabular}{l| c | c  }
& relation  &  reference \\ 
\hline
\hline
unpolarized && \\
&$F_2(x) = 2xF_1(x)$ & Callan, Gross (1969)  \cite{CG69}  
\\
\hline
\hline
polarized && \\
&$ g_2(x,Q^2) = - g_1(x,Q^2) + \int_x^1 \frac{dy}{y} g_1(y,Q^2) $&
\\
& + 
$\begin{array}{c}
{\rm twist-3\;contributions} \\ 
{\rm and\; mass\; corrections}   \\
\end{array}$
&
Wandzura,Wilczek (1977)  \cite{WW77}
\\
\hline
& $b_2(x) = 2x a_5(x) = 4x^2 a_4(x)$ & 
\\
\hline
& $2x a_1(x) =  a_2(x) +  b_1(x) +  b_2(x)$ & Dicus (1972)  \cite{Di72}
\\
\end{tabular}
\end{center}
\caption{Relations between structure functions in the Bjorken limit
         ($Q^2 \to \infty$).}
\label{tab3}
\end{table}
\begin{table}
\begin{center}
\begin{tabular}{c | c |c}
 &  parity-conserved  &  parity-violating \\
\hline
amplitude $\overline \nu + \nu$ & ${ (singlet)}\; a_S$  & ${ (valence)}\; a_V$  
\\ 
amplitude $\overline \nu - \nu$ &   $a_{\Delta V}$ & $a_{\Delta S}$ 
\\
\end{tabular}
\end{center} 
\caption{The four possible flavor combinations for neutrino scattering.}
\label{tabflav}
\end{table} 
%\newpage
%\begin{table} 
%\begin{center}
%\begin{tabular}{l| c | c | c }
%reduced matrix  & notation & theoretical value & experimental value \\
%element                & used in the reference & of sum rule calculations &           \\
%\hline \hline &&& \\ 
%$a^{(0,2)}$   &&&  \\
%                &&&  \\
%&&&\\
%$\tilde a^{(2,4)}$   & $M^2 \langle \langle O^S    \rangle \rangle /2$ &0.1867 $\pm$ 0.0560 \cite{BK87}  &    --- \\
%                & $M^2 \langle \langle O^{NS} \rangle \rangle /2$ & 0.0849 $\pm$ 0.0255 \cite{BK87}  &   --- \\
%&&&\\
%$a_5^{(0,2)}$ & $a^{(0)}_{\rm proton }$ &&  \\
%                & $a^{(0)}_{\rm neutron}$ &&  \\
%&&&\\
%$a_5^{(2,2)}$ & $a^{(2)}_{\rm proton }$ &    ---                 &  0.023 $\pm$ 0.014 \cite{Ab95}\\
%                & $a^{(2)}_{\rm neutron}$ &    ---                 & -0.007 $\pm$ 0.031 \cite{Ab95}\\
%&&&\\ 
%$\tilde a_5^{(2,3)}$ & $d^{(2)}_{\rm proton }$ &  -0.006 $\pm$ 0.003 \cite{Stein3}   &  0.0054 $\pm$ 0.0050 \cite{Ab95P}\\
%                & $d^{(2)}_{\rm neutron}$ &  -0.03  $\pm$ 0.01  \cite{Stein3}   &  0.0024 $\pm$ 0.019  \cite{Ab95P}\\
%&&&\\
%$\tilde a_5^{(2,4)}$ & $f^{(2)}_{\rm proton }$ &  -0.037 $\pm$ 0.006 \cite{Stein4}   & ---\\ 
%                & $f^{(2)}_{\rm neutron}$ &  -0.013 $\pm$ 0.006 \cite{Stein4}   & ---\\ 
%&&&\\
%\end{tabular}
%\end{center}
%\caption{Values of some reduced matrix elements: The results of sum rule calculation are compared to
%experimental values. Here always a
%summation over the charged weighted flavor indices is understood,
%e.g., $a^{(0)} = \sum_f Q_f^2 a_f^{(0)}$.}
%\label{tab4}
%\end{table}
\begin{center}
\leavevmode
\epsfxsize=11.0cm
\epsfbox{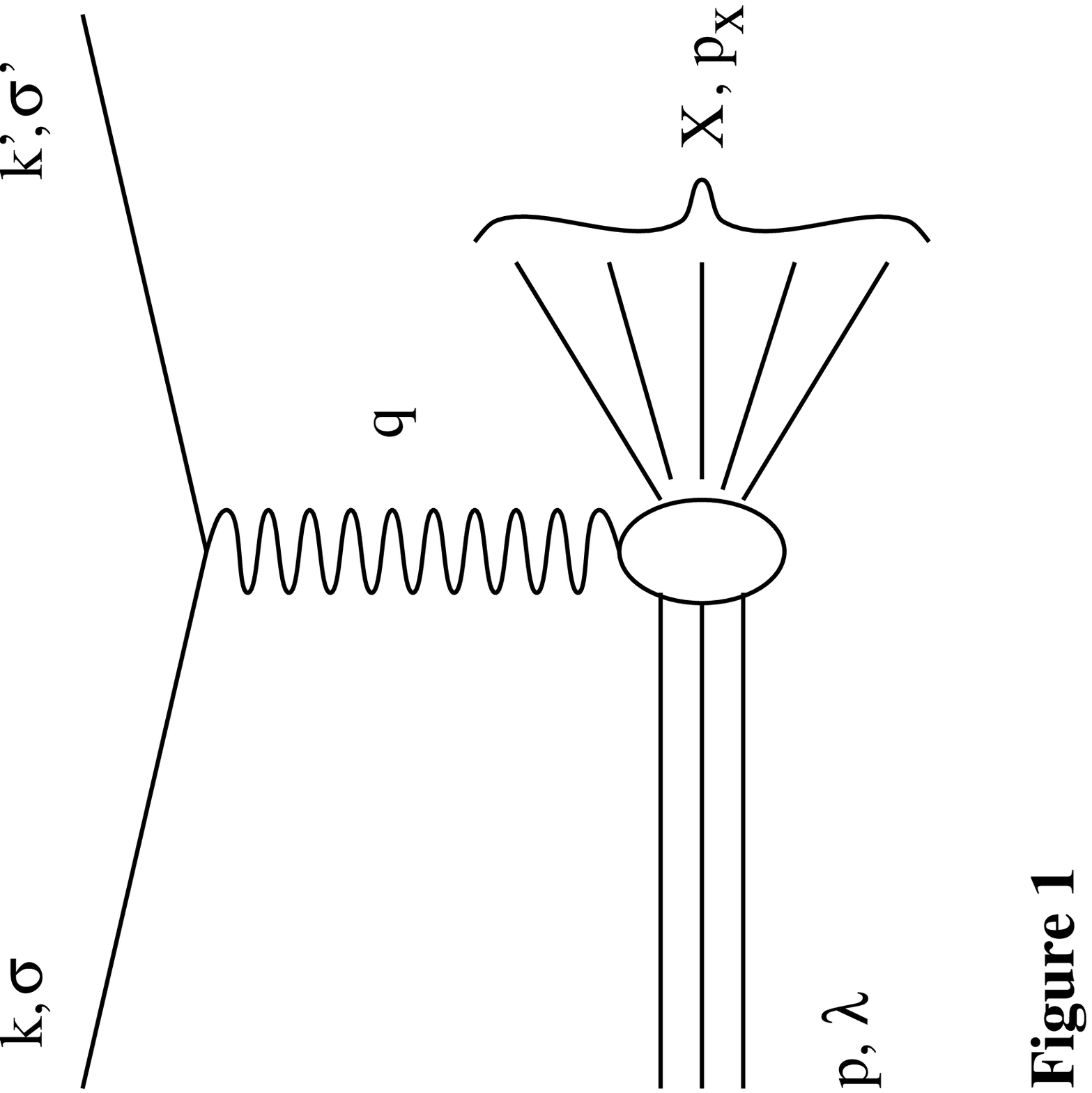}        
\newline
\newline
\end{center}
\newpage
\begin{center}
\leavevmode
\epsfxsize=11.0cm
\epsfbox{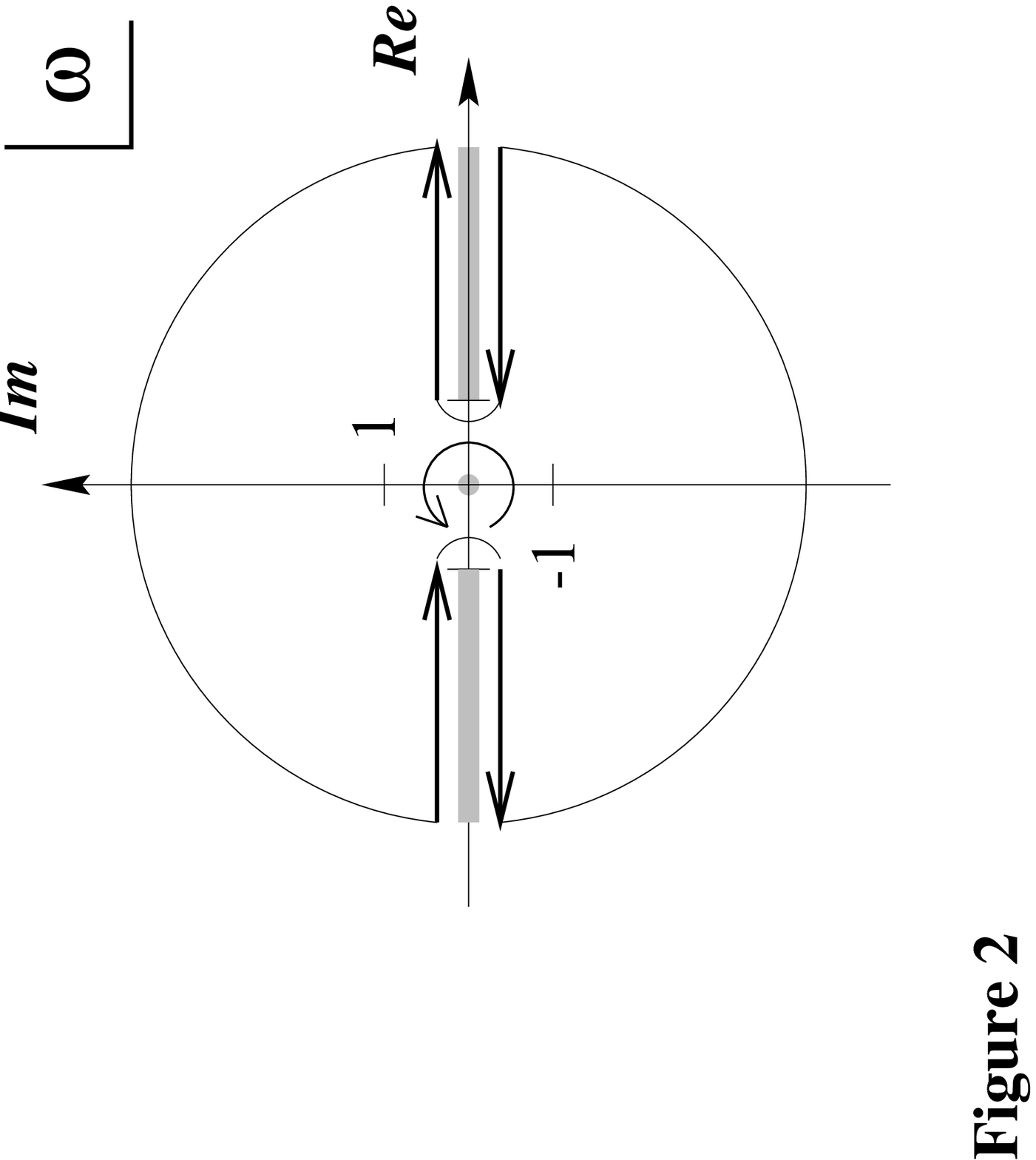}        
\newline
\newline
\end{center}
\newpage
\begin{center}
\leavevmode   
\epsfxsize=11.0cm
\epsfbox{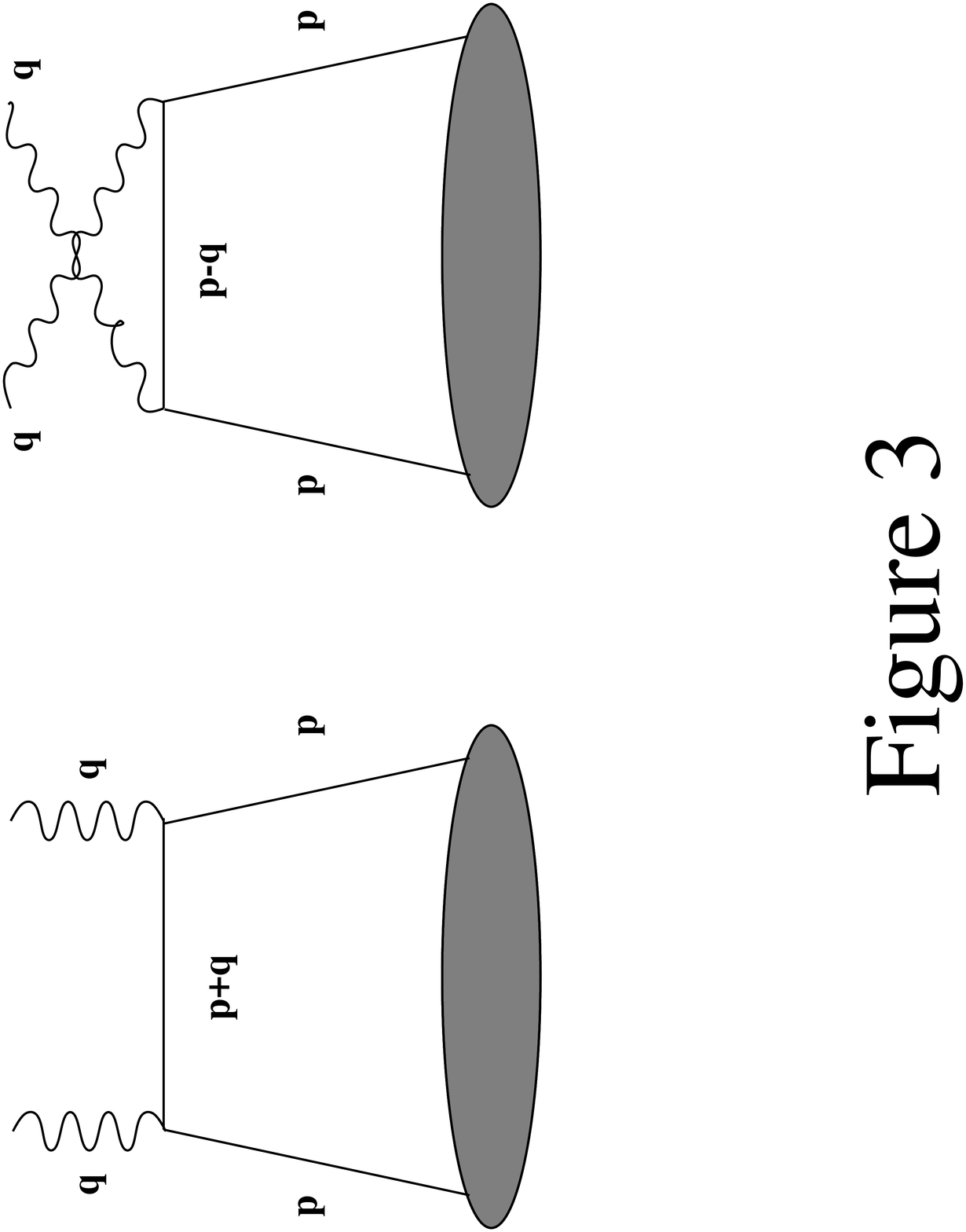}
\newline
\newline
\end{center}
\newpage
\begin{center}
\leavevmode
\epsfxsize=11.0cm
\epsfbox{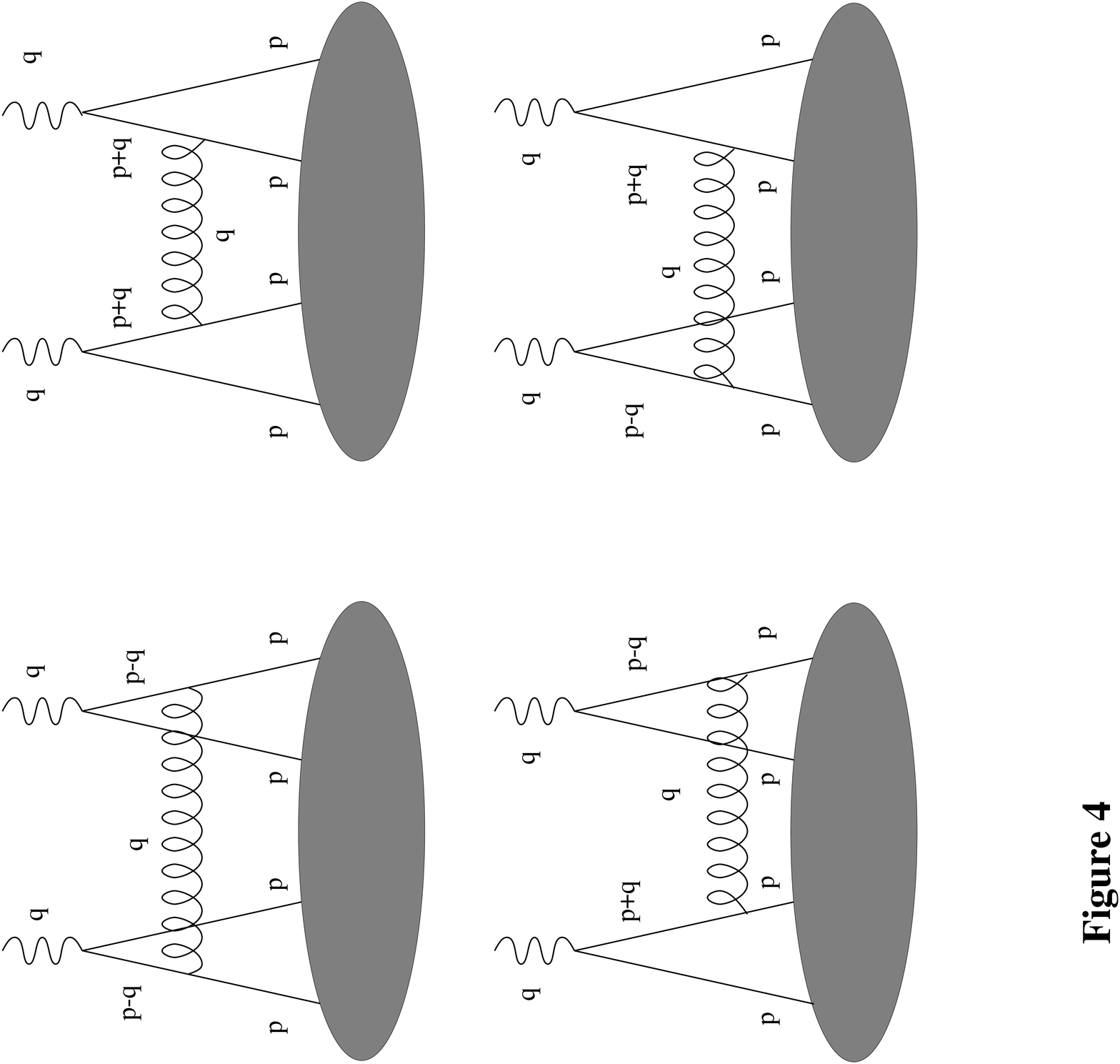}        
\newline
\newline
\end{center}
\end{document}